\documentclass[longauth]{aa}  

\usepackage{graphicx}
\usepackage{txfonts}
\usepackage{multirow}
\usepackage{amsmath}
\usepackage{xcolor}
\usepackage{ragged2e}
\newcommand{\micron}{{\mbox{$\mu$m}}}
\newcommand{\degree}{{\mbox{$^\circ$}}}
\newcommand{\mbh}{{\mbox{$M_\mathrm{BH}$}}}
\newcommand{\thetao}{{\mbox{$\theta_\mathrm{o}$}}}
\newcommand{\nodata}{{\mbox{...}}}
\newcommand{\vcirc}{{\mbox{$v_\mathrm{circ}$}}}
\newcommand{\OIII}{[O{\sevenrm\,III}]}

 \font\sevenrm=cmr7 scaled 1000

\begin{document} 


%
%
\title{The spatially resolved broad line region of IRAS~09149$-$6206}

\authorrunning{GRAVITY Collaboration}
\author{GRAVITY Collaboration:
A.~Amorim\inst{19,21} 
\and W.~Brandner\inst{22}  
\and Y.~Cl\'enet\inst{2} 
\and R.~Davies\inst{1}
\and P.~T.~de~Zeeuw\inst{1,17} 
\and J.~Dexter\inst{24,1} 
\and A.~Eckart\inst{3,18} 
\and F.~Eisenhauer\inst{1} 
\and N.M.~F\"orster~Schreiber\inst{1} 
\and F.~Gao\inst{1} 
\and P.~J.~V.~Garcia\inst{15,20,21} 
\and R.~Genzel\inst{1,4} 
\and S.~Gillessen\inst{1} 
\and D.~Gratadour\inst{2,25} 
\and S.~H\"onig\inst{5} 
\and M.~Kishimoto\inst{6} 
\and S.~Lacour\inst{2,16} 
\and D.~Lutz\inst{1} 
\and F.~Millour\inst{7}  
\and H.~Netzer\inst{8} 
\and T.~Ott\inst{1} 
\and T.~Paumard\inst{2} 
\and K.~Perraut\inst{12} 
\and G.~Perrin\inst{2} 
\and B.~M.~Peterson\inst{9,10,11} 
\and P.~O.~Petrucci\inst{12} 
\and O.~Pfuhl\inst{16}
\and M.~A.~Prieto\inst{23}  
\and D.~Rouan\inst{2} 
\and J.~Shangguan\inst{1}\thanks{Corresponding author: J. Shangguan \newline e-mail: shangguan@mpe.mpg.de}
\and T.~Shimizu\inst{1} 
\and M.~Schartmann\inst{1}
\and A.~Sternberg\inst{8,14} 
\and O.~Straub\inst{1} 
\and C.~Straubmeier\inst{3} 
\and E.~Sturm\inst{1} 
\and L.~J.~Tacconi\inst{1} 
\and K.~R.~W.~Tristram\inst{15}  
\and P.~Vermot\inst{2} 
\and S.~von~Fellenberg\inst{1}
\and I.~Waisberg\inst{13} 
\and F.~Widmann\inst{1} 
\and J.~Woillez\inst{16}}

\institute{
Max Planck Institute for Extraterrestrial Physics (MPE), Giessenbachstr.1, 85748 Garching, Germany
\and LESIA, Observatoire de Paris, Universit\'e PSL, CNRS, Sorbonne Universit\'e, Univ. Paris Diderot, Sorbonne Paris Cit\'e, 5 place Jules Janssen, 92195 Meudon, France
\and I. Institute of Physics, University of Cologne, Z\"ulpicher Stra{\ss}e 77, 50937 Cologne, Germany
\and Departments of Physics and Astronomy, Le Conte Hall, University of California, Berkeley, CA 94720, USA
\and Department of Physics and Astronomy, University of Southampton, Southampton, UK
\and Department of Physics, Kyoto Sangyo University, Kita-ku, Japan
\and Universit\'e C\^ote d'Azur, Observatoire de la C\^ote d'Azur, CNRS, Laboratoire Lagrange, Nice, France
\and School of Physics and Astronomy, Tel Aviv University, Tel Aviv 69978, Israel
\and Department of Astronomy, The Ohio State University, Columbus, OH, USA
\and Center for Cosmology and AstroParticle Physics, The Ohio State University, Columbus, OH, USA
\and Space Telescope Science Institute, Baltimore, MD, USA
\and Univ. Grenoble Alpes, CNRS, IPAG, 38000 Grenoble, France
\and Department of Particle Physics and Astrophysics, Weizmann Institute of Science, Rehovot 76100, Israel
\and Center for Computational Astrophysics, Flatiron Institute, 162 5th Ave., New York, NY 10010, USA
\and European Southern Observatory, Casilla 19001, Santiago 19, Chile
\and European Southern Observatory, Karl-Schwarzschild-Str. 2, 85748 Garching, Germany
\and Sterrewacht Leiden, Leiden University, Postbus 9513, 2300 RA Leiden, The Netherlands
\and Max Planck Institute for Radio Astronomy, Auf dem H\"ugel 69, 53121 Bonn, Germany
\and Universidade de Lisboa - Faculdade de Ci\^{e}ncias, Campo Grande, 1749-016 Lisboa, Portugal
\and Faculdade de Engenharia, Universidade do Porto, rua Dr. Roberto Frias, 4200-465 Porto, Portugal
\and CENTRA - Centro de Astrof\'isica e Gravita\c{c}\~{a}o, IST, Universidade de Lisboa, 1049-001 Lisboa, Portugal
\and Max Planck Institute for Astronomy, K\"onigstuhl 17, 69117, Heidelberg, Germany
\and Instituto de Astrof\'isica de Canarias (IAC), E-38200 La Laguna, Tenerife, Spain
\and Department of Astrophysical \& Planetary Sciences, JILA, University of Colorado, Duane Physics Bldg., 2000 Colorado Ave, Boulder, CO 80309, USA
\and Research School of Astronomy and Astrophysics, Australian National University, Canberra, ACT 2611, Australia
}

   \date{Received xxx, 2020; accepted xxx, 2020}

\abstract{We present new near-infrared VLTI/GRAVITY interferometric spectra that 
spatially resolve the broad Br$\gamma$ emission line in the nucleus of the 
active galaxy IRAS~09149$-$6206. We use these data to measure the size of the 
broad line region (BLR) and estimate the mass of the central black hole. Using 
an improved phase calibration method that reduces the differential phase 
uncertainty to 0.05\degree\ per baseline across the spectrum, we detect a 
differential phase signal that reaches a maximum of $\sim0.5\degree$ between the 
line and continuum. This represents an offset of $\sim120\,\mu$as (0.14\,pc) 
between the BLR and the centroid of the hot dust distribution traced by the 
2.3\,$\mu$m continuum. The offset is well within the dust sublimation region, 
which matches the measured $\sim0.6$\,mas (0.7\,pc) diameter of the continuum.  
A clear velocity gradient, almost perpendicular to the offset, is traced by the 
reconstructed photocentres of the spectral channels of the Br$\gamma$ line.  We 
infer the radius of the BLR to be $\sim 65\,\mu$as (0.075\,pc), which is 
consistent with the radius--luminosity relation of nearby active galactic nuclei 
derived based on the time lag of the H$\beta$ line from reverberation mapping 
campaigns.  Our dynamical modelling indicates the black hole mass is 
$\sim 1\times10^8\,M_\odot$, which is a little below, but consistent with, the 
standard $M_{\rm BH}$--$\sigma_*$ relation.}


   \maketitle
%

\section{Introduction}

Massive black holes in the centres of galaxies are a key component of galaxy 
evolution, because of the role that accreting black holes have in the feedback 
that regulates star formation and galaxy growth 
\citep{booth09,fabian12,somerville15,dubois16}.  Knowing their masses is central 
to our understanding of this co-evolution 
\citep{hopkins08,somerville08,kormendy13,heckman14}, but measuring their masses 
robustly is challenging because it requires resolving spatial scales where the 
black hole dominates the gravitational potential \citep{ferrarese05}.  For 
inactive galaxies, it can be done using spatially resolved stellar 
\citep[e.g.][]{thomas04,Onken2014ApJ,saglia16} or gas kinematics 
\citep{Hicks2008ApJS,davis14,onishi17,boizelle19}. 
These methods can be difficult to apply to active galaxies displaying broad 
emission lines because the active galactic nucleus (AGN) itself is so bright, 
although it has been possible for a few objects \citep{davies06}.  Instead, the 
most precise method for measuring black hole masses in AGNs is through megamaser 
kinematics using VLBI \citep{greene10,kuo11}. However, this requires the 
megamasers to be orbiting in a Keplerian disk restricted to near-edge-on view, 
which is very rare \citep{Zhu2011ApJ,vdBosch2016ApJ}.  Reverberation mapping 
(RM) exploits the variability of the black hole accretion to shed light on the 
size of the broad line region (BLR) and hence leads to a measurement of the 
black hole mass \citep{Blandford1982ApJ,Peterson1993PASP,Peterson2004ApJ}.  
Despite potential biases \citep{Shankar2019MNRAS} and caveats in terms of the 
virial factor $f$ that reflects the unknown geometry and kinematics of the BLR 
\citep{Onken2004ApJ,Woo2010ApJ,Ho2014ApJ}, it has been used successfully for 
many years.

Spatially resolving the very small size of the BLR, $10^3-10^5$ gravitational 
radii \citep{netzer93,Peterson1993PASP,netzer15}, has been a long-standing goal 
of spectroastrometry \citep{Petrov2001CRPhy,Marconi2003ApSS}, that has recently 
become possible with long baseline near-infrared interferometry.  GRAVITY, a 
second-generation VLTI instrument, has greatly improved both the sensitivity of 
earlier efforts, as well as combining all four of the 8-m Unit Telescope (UT) 
beams to yield six simultaneous baselines \citep{GC2017FL}. In 
\citet[][hereafter, GC18]{GC2018Natur} we reported the first robust measurements 
of BLR size and kinematics for 3C~273 by combining differential phase spectra 
with the Pa$\alpha$ emission line profile to model the BLR as a thick rotating 
disk under the gravitational influence of a black hole of 
$\sim 3\times 10^8\,\rm{M}_{\sun}$.  This value is fully consistent with the 
result of a subsequent study using 10-year RM data \citep{Zhang2019ApJ}.  We 
have now embarked on a programme to make independent estimates of the black hole 
masses in a sample of AGN. The aim is not just to verify the masses derived 
through reverberation mapping, but to understand better the structure of the 
BLR, which has implications for inflow (accretion) and outflow processes on 
small scales.

The classical picture of the BLR is a virialized distribution of clouds, with 
good evidence that many are rotating systems. This comes from a variety of 
observations including variations in the polarisation properties across the 
broad line profile \citep{smith04,smith05}. In addition, a significant minority, 
$\sim3\%$, of BLRs in both radio quiet and radio loud AGN show broad 
double-peaked profiles that are well fitted by disk emission 
(\citealt{eracleous94,eracleous03,strateva03,StorchiBergmann2017ApJ}). That 
rather few objects show such characteristics may be explained by the apparent 
relation between line width and shape, measured as FWHM/$\sigma$, reported by 
\cite{kollatschny11}.  This suggests that rapidly rotating BLRs are flattened 
systems, while slower rotating BLRs have a more spherical structure due to 
turbulence. The turbulent velocities of 300--700\,km\,s$^{-1}$ for H$\beta$ and 
2000--4000\,km\,s$^{-1}$ for C\,{\sc iv} may be indicative of outflowing gas. 
Disk winds are theoretically appealing and apply in many different situations 
including AGN \citep{emmering92,murray95,Dehghanian2020}; and an outflow at a 
characteristic elevation of 30\degree\ above the mid-plane is the basis of the 
concept proposed by \citet{elvis00}, a geometry that can explain a variety of 
observations including the broad absorption and broad emission lines. Rotating 
disk winds have been developed on a more physical basis by \citet{everett05} and 
\citet{keating12}; and PG\,1700$+$518 is one example of a rotating outflow that 
has been observed \citep{young07}. It therefore seems likely that the dynamics 
of the BLR may be a combination of rotation and outflow in a way that depends on 
the properties of individual objects.

Recently, thanks to high signal-to-noise ratio (SNR) spectra, the non-parametric 
velocity-resolved RM analyses of tens of bright AGNs have begun to reveal this 
structure, through qualitative evidence for both inflow and outflow in addition 
to virial motion \citep[e.g.][]{Denney2009ApJ,Bentz2010ApJ,Peterson2014SSRv,
Du2016ApJb,DeRosa2018ApJ,Du2018ApJa,Horne2020}.  Furthermore, temporal variation 
of the BLR geometry and dynamics has been reported for some AGNs 
\citep{Pei2017ApJ,Pancoast2018ApJ,Xiao2018ApJ}, implying that the virial factor 
of the same source may evolve with time.  

Parametric models of the BLR geometry and dynamics have been successfully 
applied to less than two dozen AGNs that have both high SNR spectra and high 
cadence monitoring 
\citep{Pancoast2014MNRASb,Grier2017ApJb,Williams2018ApJ,Li2018MNRAS}.  These 
enable one to fit for radial (inflow or outflow) motion of the clouds in 
addition to rotation, and hence also to derive the virial factors for individual 
objects.  While the number of objects is too small for robust statistics, more 
than half of the BLRs are dominated by radial motion, with inflow and outflow 
being equally common.  Remarkably, the black hole masses inferred from such 
dynamical modelling are usually consistent with those measured with the 
classical RM method, even when the BLR is dominated by outflow 
\citep{Williams2018ApJ}.  However, there are degeneracies in the models 
\citep{Grier2017ApJb}; and the impact of systematic uncertainties, such as the 
assumption that the ionisation source is point-like and that the BLR structure 
does not change significantly during the monitoring campaign, are challenging to 
assess \citep{Pancoast2014MNRASa,Grier2017ApJb}.  In particular, assumptions 
about the geometry of different BLR components may significantly bias the 
inferred physical interpretation \citep{Mangham2019MNRAS}.  As such, an 
independent method to constrain the BLR structure is much needed.

This role is fulfilled by optical/near-infrared long baseline interferometry 
combined with spectroastrometry.  In this paper we present an analysis of new 
GRAVITY observations for IRAS~09149$-$6206 ($\alpha = 09$:16:09.39, 
$\delta = -62$:19:29.9).  \citet{Perez1989AA} serendipitously discovered 
IRAS~09149$-$6206 as an AGN in the IRAS Point Source Catalogue during a search 
for planetary nebulae after optical spectra showed characteristic broad Balmer 
lines. IRAS~09149$-$6206 is at a redshift of 0.0573 \citep{Perez1989AA}, which 
means that the broad Br$\gamma$ line can still be observed in the $K$-band and 
thus GRAVITY.  Unfortunately, very little archival data exist for 
IRAS~09149$-$6206. It is detected in early radio continuum surveys, but not 
resolved \citep{Murphy2007MNRAS,Panessa2015MNRAS}.  Existing optical/NIR 
images (e.g., \citealt{Veron1990AA,Marquez1999AAS}) cannot constrain the 
morphological type of the host galaxy, and mid-infrared interferometric observations 
only barely resolve it \citep{Kishimoto2011AA,Burtscher2013AA,LopezGonzaga2016AA}. 
Section~\ref{sec:obs} describes the observations and data reduction including 
changes which significantly improve the precision of our differential phase 
spectra.  In Section~\ref{sec:blrpos}, we show that the model-independent 
photocentre positions reveal the detection of a spatial offset between the hot 
dust continuum and Br$\gamma$, and a velocity gradient across the emission line.  
Building on this, in Section~\ref{sec:rot_out} we adopt a kinematic cloud model 
to fit the phase data in order to constrain the BLR size and kinematics.  In 
Section~\ref{sec:bhmass}, we discuss the implications of our measured 
$M_{\rm BH}$ and place IRAS~09149$-$6206 on the BLR radius--luminosity relation.  
The possible interpretations of the offset between the BLR and the continuum is 
discussed in Section~\ref{sec:offset}.

This work adopts the following parameters for a $\Lambda$CDM cosmology: 
$\Omega_m = 0.308$, $\Omega_\Lambda = 0.692$, 
and $H_{0}=67.8$ km s$^{-1}$ Mpc$^{-1}$ \citep{Planck2016AA}. Using this 
cosmology, 1~pc subtends 0.87~mas on sky and 1~$\mu$as corresponds to 1.37 light 
day at the redshift of IRAS~09149$-$6206.

\section{Observations and Data Reduction} \label{sec:obs}

\begin{table}
\centering
\begin{tabular}{l|c|c|r@{$-$}l}
\hline\hline
Date        & On-source      & Seeing        & \multicolumn{2}{c}{Coherence} \\
            & time (min)     & (\arcsec)     & \multicolumn{2}{c}{time (ms)} \\ \hline
2018 Nov 20$^\ast$    & 55             & 0.65 $-$ 0.97 &  3.3 & 4.2 \\
2019 Feb 16$^\dagger$ & 75             & 0.53 $-$ 0.95 &  6.0 & 9.1 \\
2019 Nov 07$^\ast$    & 85             & 0.36 $-$ 0.71 &  6.5 & 10.7 \\
2019 Nov 09$^\ast$    & 30             & 0.42 $-$ 0.58 &  2.4 & 4.2 \\
2019 Dec 09$^\ast$    & 45             & 0.50 $-$ 0.71 &  1.8 & 3.9 \\
2020 Feb 09           & 55             & 0.47 $-$ 0.69 & 13.2 & 15.4 \\
2020 Feb 10$^\dagger$ & 75             & 0.51 $-$ 0.99 &  7.7 & 13.2 \\
2020 Mar 08$^\dagger$ & 45             & 0.49 $-$ 0.80 &  7.1 & 9.9 \\
\hline\hline
\end{tabular}
\caption{Exposure time and weather conditions for the observations.
The reported data only include exposures for which we were able to 
track the fringe for $>80\%$ of the exposure time.  The seeing and coherence 
time are based on measurements from the Differential Image Motion Monitor and 
Multi-aperture Scintillation Sensor on Paranal. \newline
$^\ast$ K stars are observed as calibrators. \newline
$^\dagger$ B stars are observed as calibrators.}
\label{tab:obs}
\end{table}

We observed IRAS~09149$-$6206 with GRAVITY \citep{GC2017FL} using all four UTs, 
on eight occasions between November 2018 and March 2020, primarily as part of a 
Large Programme to spatially resolve the BLR and measure black hole masses in a 
sample of AGN.\footnote{Observations were made using the ESO Telescopes at the 
La Silla Paranal Observatory, program IDs 0102.B-0667 and 1103.B-0626.} 
Targets were selected as the brightest type~1 AGNs visible 
from the VLTI and above the GRAVITY sensitivity limit ($K<11$).  
IRAS~09149$-$6206, in particular, has more than 70\% of its total $K$ band flux originating in the 
nucleus and is also bright and compact in the optical ($R \approx 11.5$).  These properties 
make it an ideal GRAVITY target for observations where the source is phase referenced to itself as well as used for adaptive optics.
We adopted the single-field on-axis mode for all of the observations with combined polarisation.  The scientific spectra 
(1.95--2.45~\micron) were taken in medium-resolution mode 
($\lambda/\Delta \lambda \approx 500$), with 90 independent spectral elements, 
extracted and resampled into 210 channels. 

All observations followed a similar sequence.  After the telescope had pointed 
to the target and the adaptive optics (MACAO; \citealt{Arsenault2003SPIE}) had 
closed the loop, the telescope beams were coarsely aligned on the VLTI 
laboratory camera IRIS \citep{Gitton2004SPIE}.  The light was then fed to 
GRAVITY and the internal beam tracking of GRAVITY aligned the fringe-tracking 
(FT) and science channel (SC) fibres on the target.  After the fringe tracker 
had searched for and found the fringe, the scientific exposures were started. 
Each set of scientific exposures consisted of ten frames of 30-s integration 
(NDIT=10 and DIT=30~s).  Fringe tracking is difficult for faint targets and 
leads to large phase noise ($\gtrsim 0.5\degree$).  We therefore integrated 
deeply on-source, with only occasional sky exposures.  In order to account for 
the observatory transfer function (e.g. coherence loss due to vibrations, 
uncorrected atmosphere, birefringence, etc.), we observed a calibrator star 
close to the science target.  The calibrator data are used to calibrate the 
complex visibility and closure phase of the FT data, as well as the flux 
spectrum of the AGN.  A calibrator was observed for all runs except 2020 Feb 09.  
The date, exposure time, and weather conditions for all observations are 
summarised in Table~\ref{tab:obs}.

GRAVITY measures the complex visibilities of six baselines (telescope 
pairs).  The visibility amplitudes measure the angular extent of a structure, 
whereas the phases provide its position and spatial distribution on the sky 
(e.g., \citealt{Buscher2015}).  The GRAVITY synthesis beam (about 3~mas) is usually 
much larger than the BLR of an AGN at cosmological distance and can only 
marginally resolve the continuum emission from the hot dust 
\citep{GC2018Natur,GC2020HD}.  Therefore, the differential phase and 
differential visibility amplitude are the most important interferometric 
observables for the BLR.  Spectro-astrometry \citep{Bailey1998SPIE} enables us to spatially resolve the BLR, using the 
continuum as a reference.  The differential phase measures the shift in the 
photocentre on sky at different wavelength channels of the broad line 
emission with respect to the continuum (see Appendix~\ref{apd:visphase}).  The 
differential visibility amplitude measures the relative size difference between 
the broad line emission and the continuum emission (see 
Section~\ref{sec:blrvamp}).

\subsection{Pipeline data reduction} \label{ssec:pip}

We used the latest version of the GRAVITY pipeline to reduce the data 
\citep{GC2017FL,Lapeyrere2014SPIE}.  This includes the recent improvements 
to lower the imprint of the internal source of the calibration unit 
\citep{Blind2014SPIE}, and a better filtering of hot pixels due to cosmic rays. 
For the FT continuum visibility data, we used the default pipeline settings. 
However, the low SNR of the FT data and loss of coherence during scientific 
exposures often caused the pipeline to flag most of the DITs in one scientific 
exposure.  In analysing these data, we found a substantial improvement in 
residual phase noise (rms scatter) by retaining all data independent of FT SNR 
or the estimated visibility loss\footnote{We set the thresholds of the pipeline, 
\texttt{snr-min-ft} and \texttt{vfactor-min-sc}, to zero.} 
(GC18; \citealt{GC2020HD}, hereafter GC20a).

The pipeline reduces the data using a pixel-to-visibility matrix 
\citep{Lapeyrere2014SPIE,Lacour2019AA}.  This matrix, encoding the relative 
throughput, coherence, phase shift, and cross-talk for each pixel, was measured 
during the day with the GRAVITY calibration unit after every night that GRAVITY 
observations were performed. Applying the matrix to the detector frames yields 
the instrument-calibrated complex visibilities. The pipeline then fits the phase 
of the science channel in each frame with a 3rd-order polynomial to derive the 
phase reference, which is then subtracted from the phase. This yields the 
self-referenced complex visibility for every frame. Before averaging the complex 
visibilities of an individual exposure, we applied the algorithm developed for 
VLTI/AMBER \citep{Millour2008SPIE} to calculate and remove the average phase 
using all of the other wavelength channels for each channel.  This method 
produces consistent results and improves our phase errors, typically by about 
10\%--20\%.  Initial uncertainties for the combined visibilities are estimated 
by bootstrapping the different frames in the pipeline.  However, this 
underestimates the uncertainty, so we adopt a different method to estimate the 
phase uncertainty after the pipeline data reduction.

\begin{figure}
\centering
\includegraphics[width=0.4\textwidth]{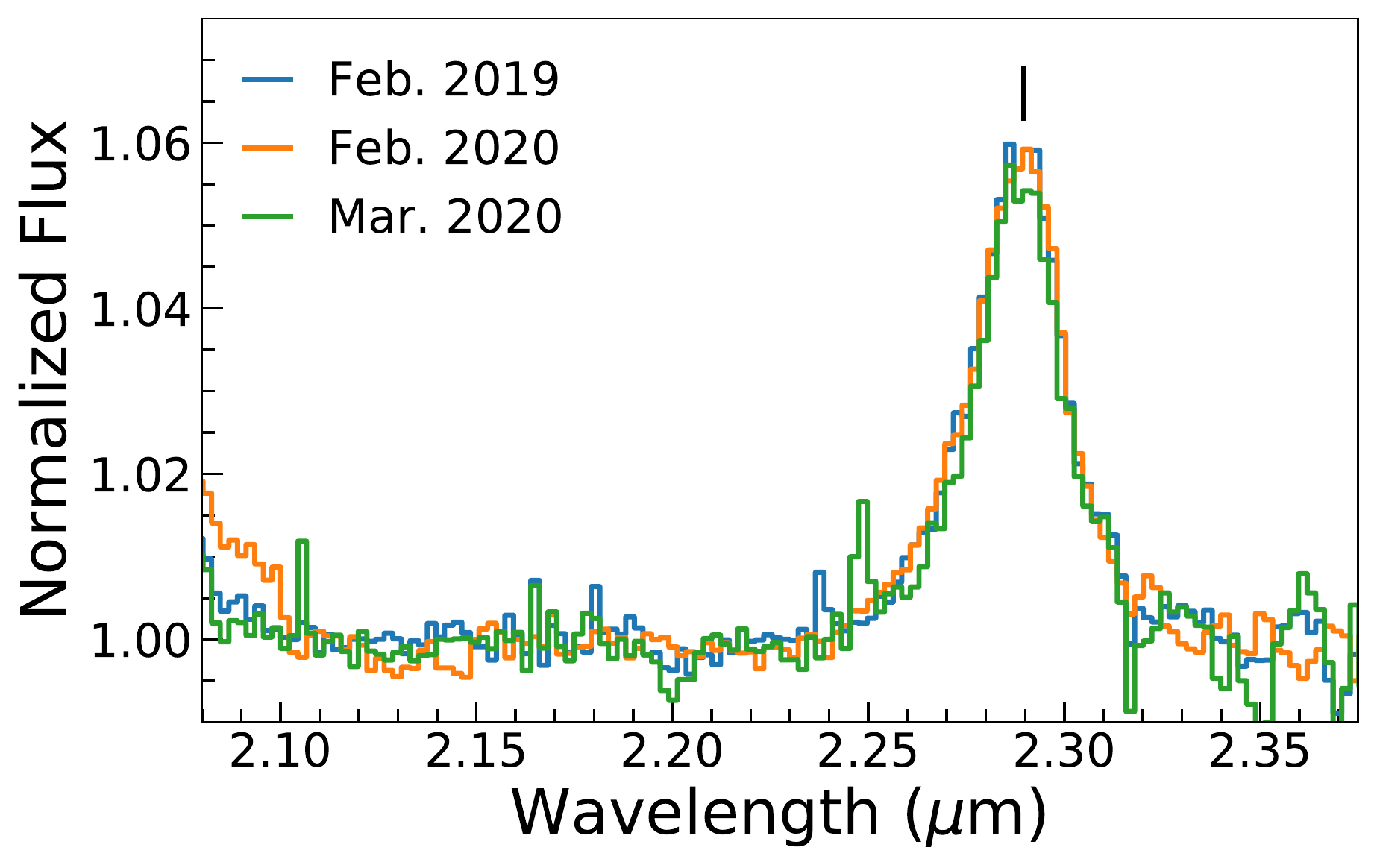}
\caption{AGN flux spectra of IRAS~09149$-$6206 normalised to the continuum. The 
black vertical bar at 2.2896~\micron\ marks the wavelength of Br$\gamma$ given a 
redshift $z=0.0573$ \citep{Perez1989AA}.  We discard the wavelength ranges 
($\lesssim 2.10\,\micron$ and $\gtrsim 2.35\,\micron$) that are significantly 
affected by the atmospheric absorption.}
\label{fig:spec}
\end{figure}

\subsection{Normalised profile of the broad Br$\gamma$ line} \label{ssec:spec}

The emission line profile, normalised to the continuum, is necessary to derive 
the velocity gradient and model the dynamics of the BLR. However, it is 
challenging to derive the Br$\gamma$ line profile for IRAS 09149$-$6206, because 
the line is red-shifted to $\sim$2.2896~\micron\ where atmospheric water 
absorption severely affects the red wing. Because late-type stars are usually 
selected as interferometric calibrators to correct for these telluric features, 
stellar absorption lines around 2.3~\micron\ complicate the correction despite 
the use of stellar templates.  To address this, we observed B star calibrators 
for three nights (different stars in different nights; Table~\ref{tab:obs}); and 
by using the flux spectra from only these nights we were able to accurately 
recover the Br$\gamma$ line profile.  Fig.~\ref{fig:spec} shows that the line 
profiles from the three nights with early-type calibrators are consistent with 
each other.  We averaged these line profiles, weighted by their statistical 
uncertainties, to derive the final line profile.  The averaged line profile is 
displayed in both Fig.~\ref{fig:phs} and Fig.~\ref{fig:vamp} to guide the eye, 
as both the differential phase and differential visibility amplitude scale 
with the normalised line profile.  We do not observe a distinct narrow emission 
line component because it is weak and its width \citep{Perez1989AA} is 
comparable to our 0.002~\micron\ channel resolution.

The standard deviation of the three spectra around the line region is about 
0.002 (normalised units), corresponding to $\sim$5\% (i.e., 0.002/0.04) of the 
line core region.  This is consistent with the GRAVITY flux uncertainty we 
measured for 3C 273 (GC18). It is larger than the statistical uncertainties 
propagated from the uncertainty of each individual exposure spectrum, which is 
the result of stacking many exposures together in each night. The systematic 
uncertainty mainly arises from the variation of the sky absorption and the 
calibrator data.  Therefore, we quadrature summed the systematic uncertainty 
(the RMS) and the statistical uncertainty of the normalised spectral flux in 
each wavelength channel with a minimum systematic uncertainty of 0.002. The 
resulting total flux uncertainty is largely uniform across the whole line 
region.  We note that our results are not significantly different if we adopt a 
slightly different threshold (e.g., 0.0015--0.0025).

\begin{figure}
\centering
\includegraphics[width=0.4\textwidth]{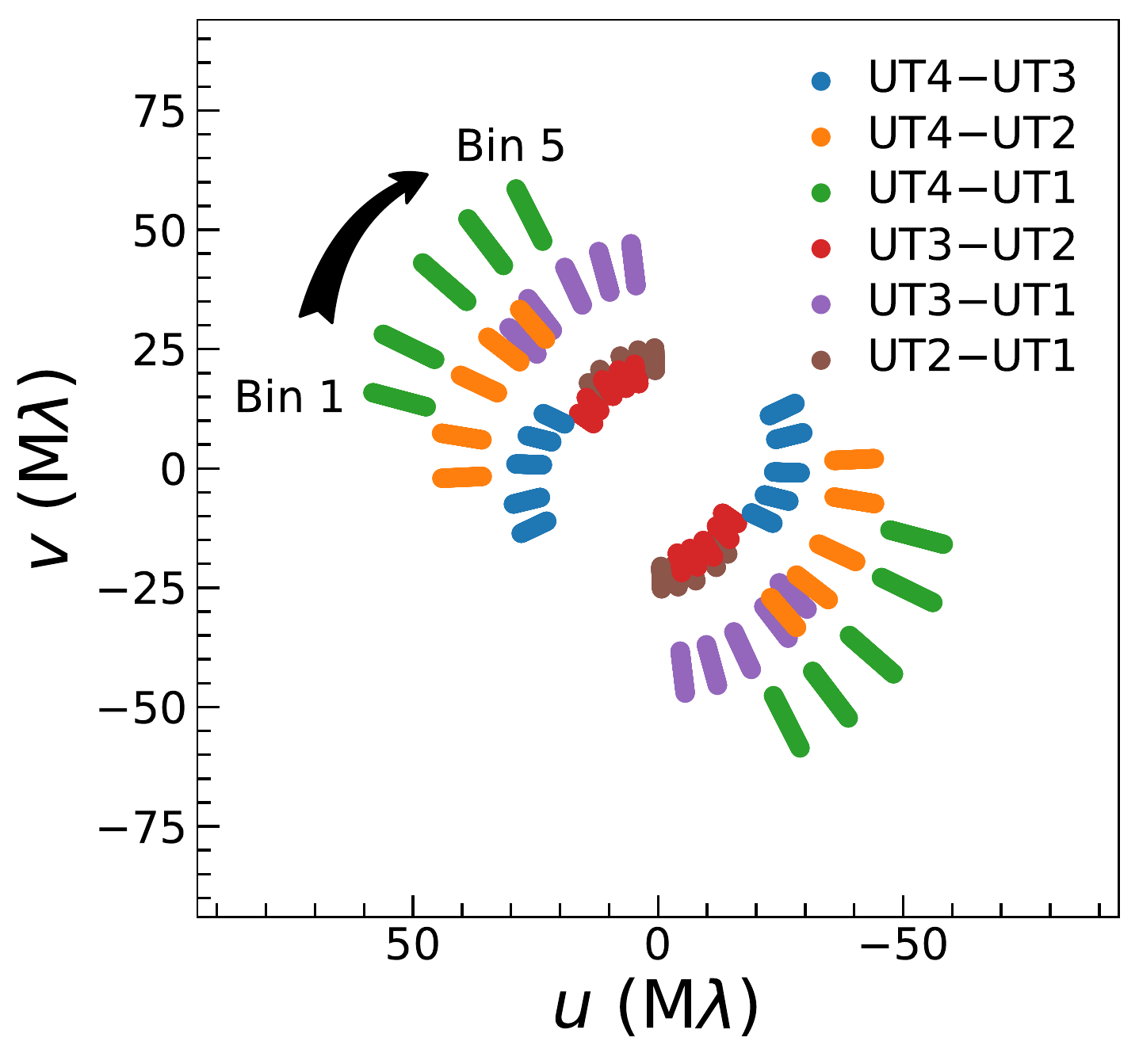}
\caption{$uv$ coordinates of the binned data. Each coloured stripe spans the 
averaged $uv$ coverage of a given baseline over the spectral range.  From bin~1 
to bin~5, the $uv$ coverage of the 6 baselines rotates clockwise.}
\label{fig:uvc}
\end{figure}

\begin{figure*}
\centering
\includegraphics[width=0.9\textwidth]{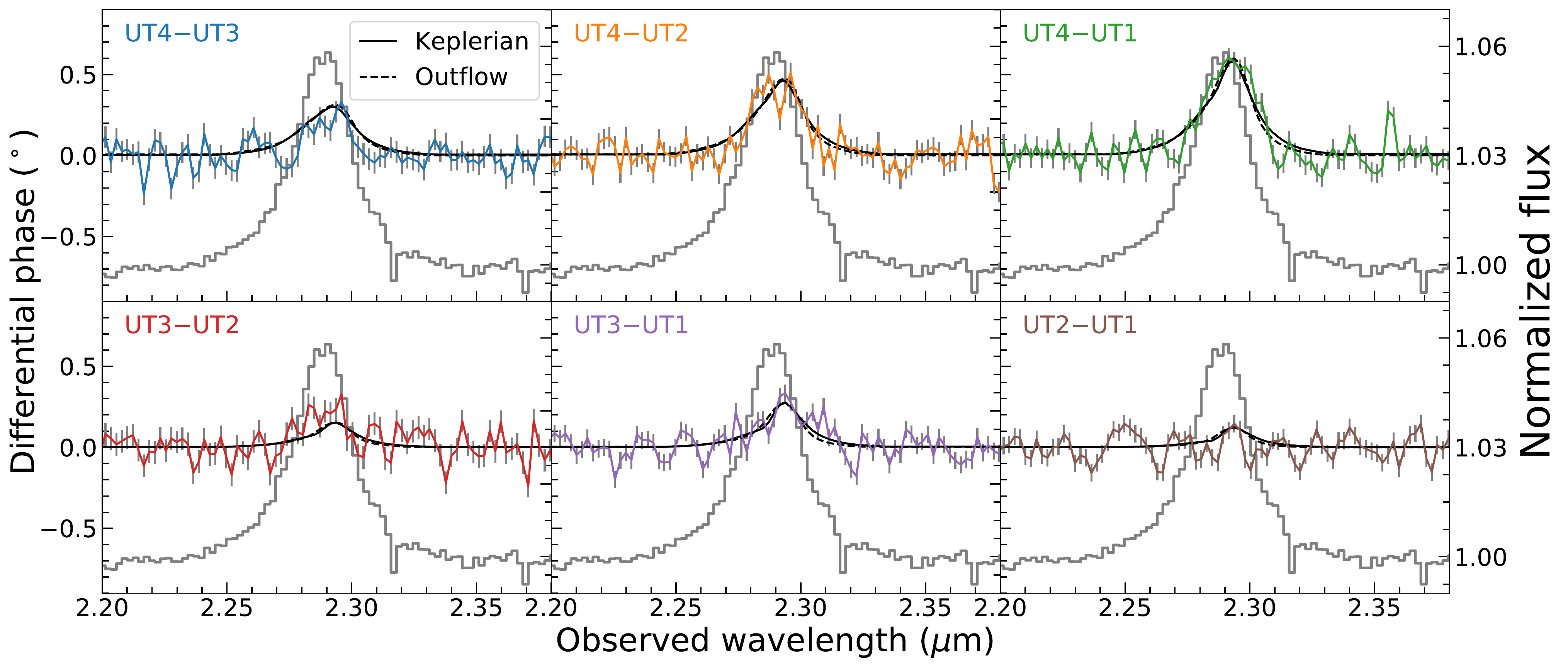}
\caption{Averaged differential phase spectra (coloured) and the normalised 
Br$\gamma$ spectrum (grey histogram) for IRAS~09149$-$6206. Solid and dashed 
black curves show the differential phase spectra of the best-fit BLR models (see 
Sec.~\ref{sec:rot_out}). For the purpose of presentation, we average all of the 
data in each baseline. BLR modelling, however, was done on the five bins based 
on $uv$ coverage as shown in Fig.~\ref{fig:uvc}.  The same Br$\gamma$ 
spectrum is displayed in all of the panels.
}
\label{fig:phs}
\end{figure*}

\subsection{The averaged differential phase} \label{ssec:vphi}

The phase of the pipeline reduced visibility shows significant 
($\gtrsim 1\degree$) instrumental features across the spectral band, which also 
varies between exposures.  As discussed in Appendix~\ref{apd:phase}, we find 
that the instrumental features primarily consist of two components.  A variable 
component is introduced by the dispersion of the air in the non-vacuum delay 
line of the VLTI, while a stable component is likely generated inside the 
cryostat of GRAVITY. Crucially, the variable component has a stable profile as a 
function of wavelength and only the total amplitude changes with each exposure. 
We remove these systematic features by fitting and subtracting a simple 
instrumental phase model -- a stable component plus a fixed phase profile with a 
variable amplitude.  Using calibrator star data, we demonstrate that the 
systematic uncertainty of our flattening method reaches $\lesssim 0.05$\degree\ 
across most of the wavelength coverage (2.05--2.40 \micron) of GRAVITY.

When we remove the residual instrumental features described above, we also mask 
out the wavelength range (2.25--2.35 \micron) where the broad Br$\gamma$ line in 
IRAS~09149$-$6206 dominates.  The mask does not affect the flattening (see 
Appendix~\ref{apd:phase}). Similarly, the uncertainty of the differential phase 
is estimated from the RMS of the flattened phase at 2.05--2.40~\micron, masking 
the wavelength range of the broad Br$\gamma$.  In addition, we also fit a 2nd 
order polynomial function locally around the Br$\gamma$ line, avoiding the line 
core (2.20--2.27 \micron\ and 2.31--2.38~\micron), to further flatten the phase 
around the line.  This is necessary because IRAS 09149$-$6206 shows a slight 
positive phase gradient across the broad emission lines, including Br$\gamma$ 
(2.2896~\micron), Br$\delta$ (2.0551~\micron), and Pa$\alpha$ 
(1.9821~\micron).\footnote{Our spectrum only covers the red wing of the 
Pa$\alpha$ line.  The overlapping line profile and phase signal of Br$\delta$ 
and Pa$\alpha$ are significantly affected by atmospheric carbon dioxide 
absorption at $\sim 2\,\micron$.  Therefore, it is very difficult to incorporate 
the Br$\delta$ line in the analysis.} Without this step, the pipeline would 
create a phase dip around the blue wing of the Br$\gamma$ line because it 
derives the phase reference without considering the scientific signal. Using the 
calibrator data, we find that the additional flattening procedure does not 
increase the noise with respect to the systematic uncertainty 
($\sim 0.05\degree$; Appendix~\ref{apd:phase}).  Finally, as shown in 
Fig.~\ref{fig:uvc}, we split the data collected across the three years into 5 
angular $uv$ bins in order to minimise smearing of the phase signal. We then 
averaged the differential phase in each spectral bin, weighted by the 
uncertainty, resulting in total integration times for each bin between 1.3 and 
1.8 hours. The full set of binned differential phase spectra can be seen in 
Appendix~\ref{apd:fit} while in Fig.~\ref{fig:phs} we show the averaged 
differential phase spectra for each baseline.  The six baselines have 
different orientations and $uv$ distances (Fig.~\ref{fig:uvc}).  The longer 
baselines are more sensitive to the small scale signal.  Therefore, the signals 
measured by different baselines behave differently.

\begin{figure*}
\centering
\includegraphics[width=0.9\textwidth]{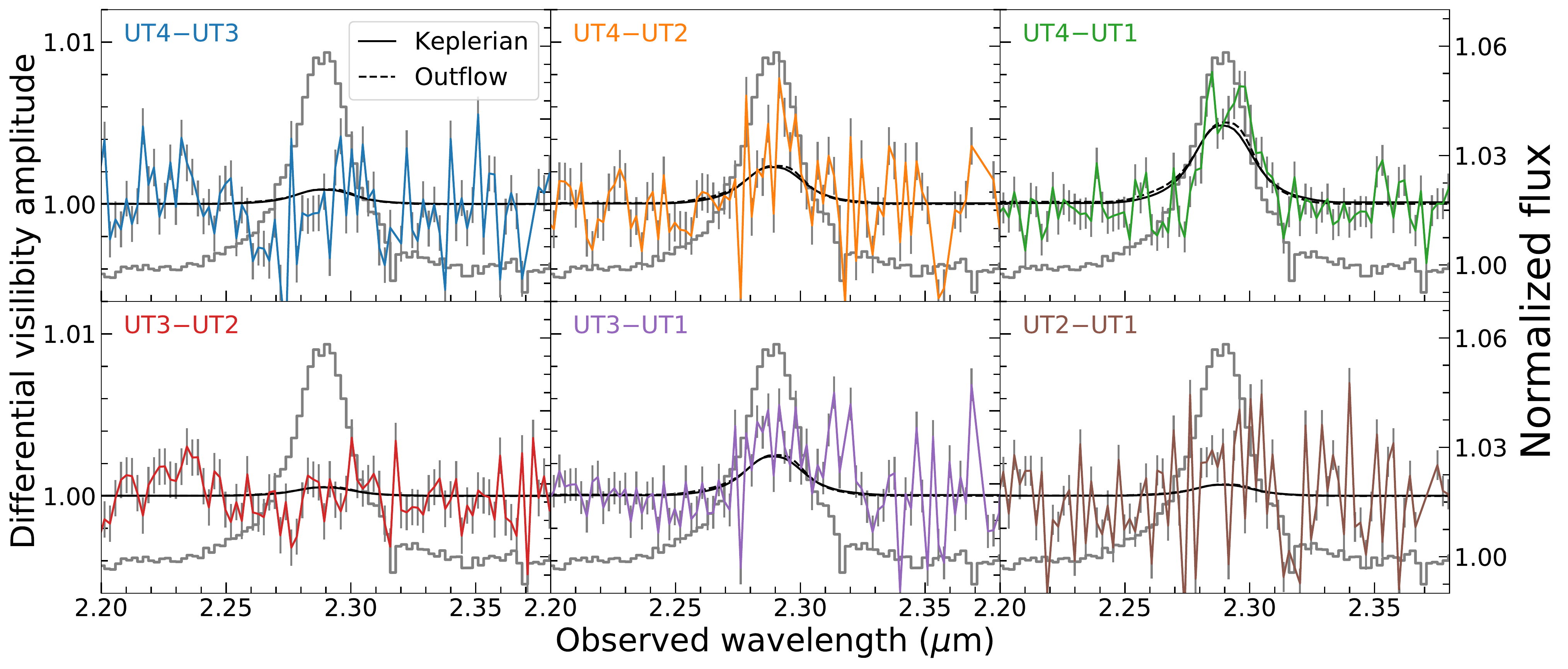}
\caption{Averaged differential visibility amplitude spectra (coloured) and the 
normalised Br$\gamma$ spectrum (grey histogram) for IRAS~09149$-$6206.  There is 
a clear positive differential visibility amplitude signal in the UT4$-$UT1 
spectrum, which follows well the line profile. Solid and dashed black curves 
show the differential visibility amplitude of the best-fit BLR models (see 
Sec.~\ref{sec:blrvamp}).  Several channels with large scatter at 
$>2.3\,\micron$ are masked for clarity.  The same Br$\gamma$ spectrum is 
displayed in all of the panels.
}
\label{fig:vamp}
\end{figure*}

\subsection{The differential visibility amplitude} \label{ssec:vamp}

To produce the differential visibility amplitude, we fit and divide out a 2nd 
order polynomial from the pipeline reduced SC visibility amplitude while masking 
the Br$\gamma$ line region.  The uncertainty is estimated from the RMS of the 
resulting normalised continuum channels.  We then average the differential 
amplitudes from each exposure together, weighted by their inverse variance, and 
the final uncertainty is calculated by propagating the individual exposure 
uncertainties. Instead of $uv$ binning, we simply average all exposures for a 
single baseline together, since the signal in the differential visibility 
amplitude is much weaker than the differential phase.  As shown in 
Fig.~\ref{fig:vamp}, the differential visibility amplitude of UT4$-$UT1 is well 
above 1 and follows the profile of the emission line.  The small dip in the 
differential visibility amplitude signal around the central wavelength of the 
line may be due to the narrow Br$\gamma$ component; but it does not affect our 
analysis because it is barely significant compared to the noise level.

\begin{figure}
\centering
\includegraphics[width=0.4\textwidth]{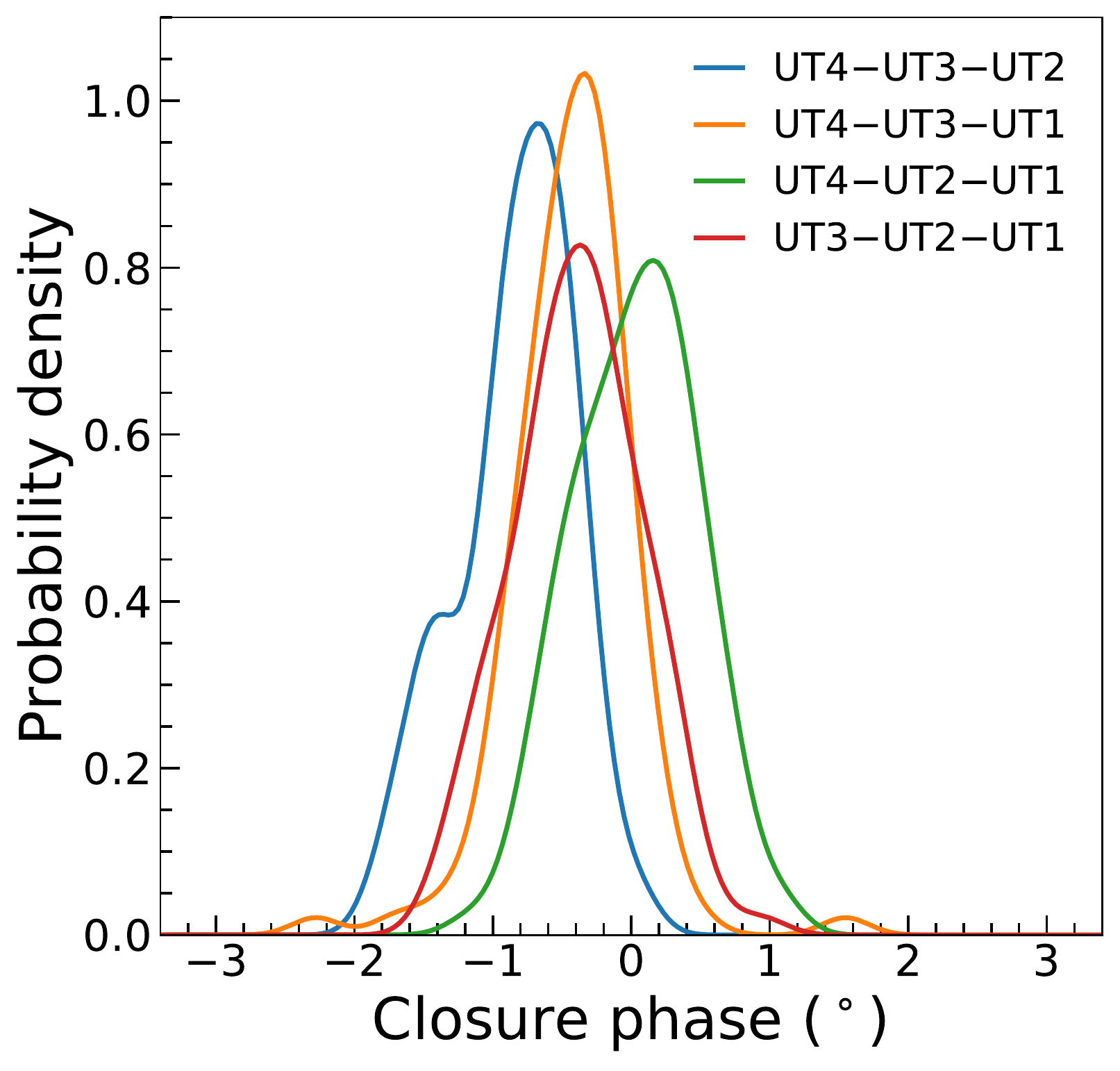}
\caption{Kernel density estimates for the closure phase distribution from the FT 
data of IRAS~09149$-$6206. The closure phase distributions for the four 
triangles are consistent with 0\degree.}
\label{fig:t3}
\end{figure}

\subsection{Continuum visibility from the fringe tracker} 
\label{ssec:ft}

The fringe tracker kept a record of fringe measurements at 300~Hz throughout 
the exposures.  There are six spectral channels across $K$ band.  The visibility 
amplitude and closure phases are important to constrain the continuum emission.  
In Fig.~\ref{fig:t3}, we plot the closure phase distribution for the four 
triangles. All distributions are consistent with an average closure phase 
between $-1\degree$ and 0\degree.


\begin{figure*}
\centering
\includegraphics[width=0.9\textwidth]{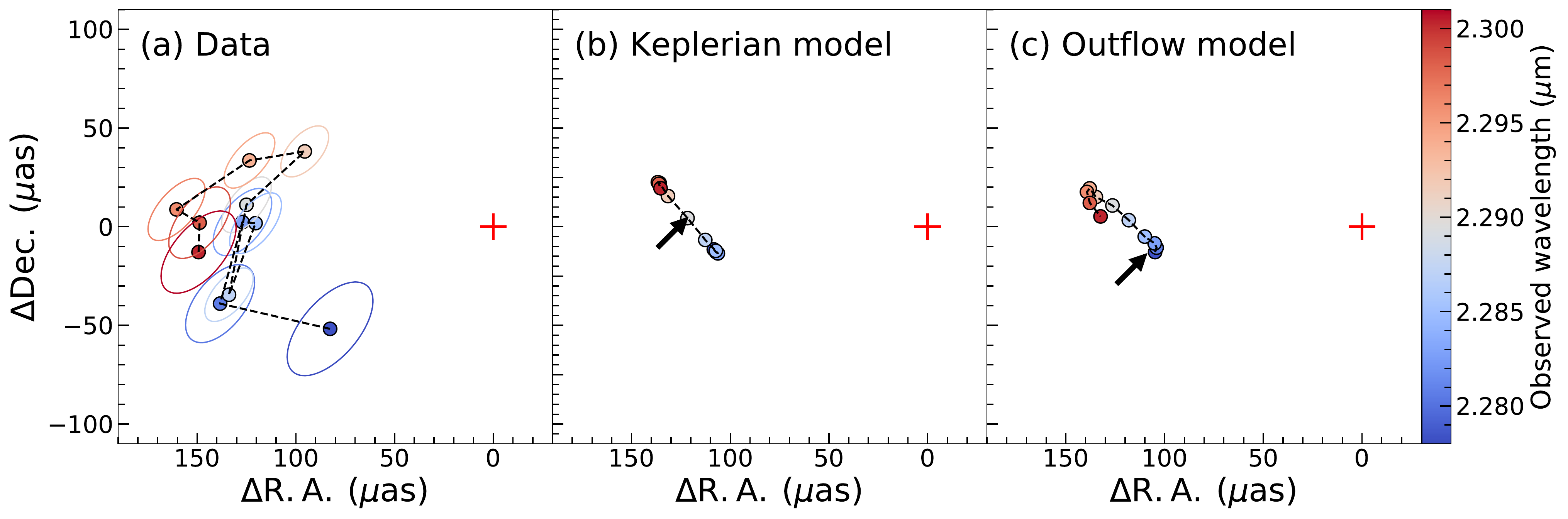}
\caption{Best-fit centroids to the differential phases of (a) observed data of 
IRAS~09149$-$6206, (b) the best-fit Keplerian model described in 
Sec.~\ref{sec:rotmod}, and (c) the best-fit outflow model described in 
Sec.~\ref{sec:outmod}.  The colour code represent the wavelength of the channels 
around the centre of the Br$\gamma$ line.  The coloured ellipses around each 
centroid in panel (a) represents the 68\% (1~$\sigma$) credible intervals of the 
uncertainty.  The red plus sign at the origin represents the photocentre of the 
continuum.  The black arrows in panel (b) and (c) indicate the origin of the BLR 
according to the inferred offset of the models.
}  
\label{fig:phc}
\end{figure*}

\section{Locating the Broad Line Region}
\label{sec:blrpos}

The differential visibility amplitude spectra provide direct evidence that the 
BLR has been unambiguously detected. The signal is especially clear in the 
UT4$-$UT1 baseline shown in Fig.~\ref{fig:vamp}, where the differential 
visibility amplitude significantly increases across the BLR line profile.  
The channels dominated by broad Br$\gamma$ emission display higher 
visibility amplitude, hence smaller size, than those dominated by the continuum 
(see Sec.~\ref{sec:blrvamp}).  Consistent with GC20a, this indicates that the 
BLR is more compact than the near-infrared continuum, which traces the hot dust 
distribution around the AGN.  IRAS 09149$-$6206 also shows a strong differential 
phase signal primarily in the UT4$-$UT2 and UT4$-$UT1 baselines. Remarkably, as 
is apparent in Fig.~\ref{fig:phs}, the signal is also entirely positive and 
peaks near the line centre, which is different from the `S-shape' seen in 3C~273 
(GC18) that crosses zero at the line centre.

A differential phase following the shape of the line profile is expected for a 
constant phase difference between the hot dust continuum and the Br$\gamma$ 
emission.  Both sources are marginally resolved, which is strongly supported by 
the $\sim 0\degree$ closure phase in Fig.~\ref{fig:t3}.  This means that to first 
order a phase difference measures an offset in photocentre position.  The phase 
signal caused by this offset, we hereafter refer to as the `continuum phase'.  
By construction, the differential phase data are referenced to the photocentre 
position of the hot dust continuum. Hence, we have the differential phase, 
$\Delta \phi_\lambda = -2\pi\, [f_\lambda/(1+f_\lambda)]\,\vec{u} \cdot 
\vec{x}_{\mathrm{BLR},\lambda}$, where $f_\lambda$ is the line flux at 
wavelength $\lambda$ relative to a continuum level of unity, $\vec{u}$ is the 
$uv$ coordinate of the baseline, and $\vec{x}_{\mathrm{BLR},\lambda}$ is the 
coordinate of the photocentre w.r.t. the centroid of the continuum (see 
Appendix~\ref{apd:visphase} for details).  Fitting the photocentre 
coordinate of each channel to the differential phase data of 30 baselines (6 
baselines $\times$ 5 angular bins), we can reconstruct the photocentres of 
the IRAS~09149$-$6206 Br$\gamma$ line emission. 

The result in Fig.~\ref{fig:phc}\,(a) shows a systematic offset of the BLR 
photocentres from the origin by $\sim 120\,\mu$as to the east.  Moreover, there 
is clear evidence for a velocity gradient that is nearly perpendicular to the 
offset: the blueshifted channels lie predominantly to the south of the 
red-shifted channels.  While the separation between individual channels is only 
moderate given the uncertainties, the general gradient from North to South 
appears robust.  Following GC18, we estimate the significance of the offset 
between the blue and red channels with an F-test, comparing the null hypothesis 
that the phase signal is produced by a single position of the unresolved BLR 
with the hypothesis that the blue and red channels are at two distinct 
positions.  Using the same channels as shown in Fig.~\ref{fig:phc}\,(a) (6 blue 
channels $<2.2896\,\micron$ and 5 red channels $>2.2896\,\micron$), we find the 
red-blue photocentre offset is at $>8\,\sigma$ significance.  If we use only 
alternate channels, in order to avoid the impact of possible data correlation, 
the same method still yields a significance $>5\,\sigma$.\footnote{The result 
remains robust independent of whether one includes the bluest channel that is 
furthest offset from the photocentres of the other channels.} According to this 
simple model, the average photocentre displacement on sky is 
$33 \pm 8 \, \mu$as, which can be considered as the lower limit of the 
separation of the photocentres.

This red-blue photocentre displacement, while model independent, is only a lower 
limit to the true physical BLR size.  To measure the physical size and constrain 
the BLR kinematics, a model is required.  We adopt a flexible model, with a full 
differential phase (see also Appendix \ref{apd:visphase}),
\begin{equation}
\label{eq:dphi_fit_text}
\Delta \phi_\lambda = [f_\lambda/(1+f_\lambda)]\, (\phi_{\mathrm{BLR},\lambda} - 
2\pi\, \vec{u} \cdot \vec{x}_\mathrm{o}),
\end{equation}
where $\vec{x}_\mathrm{o}$ is the coordinate of the origin of the BLR with 
respect to the centroid of the continuum. This velocity-independent photocentre 
offset could for example result from asymmetric structure in the continuum, or a 
physical offset between the BLR and hot dust continuum. The kinematic model 
described in the next section provides the velocity-dependent phase 
$\phi_{\mathrm{BLR},\lambda}$ of the BLR itself.

\section{Rotation versus Outflow in the Broad Line Region} 
\label{sec:rot_out}

The GRAVITY measurements of the line and phase profile for 3C~273 were modelled 
with a simple model consisting of a symmetric distribution of clouds in 
circular orbits, which fitted those data very well GC20a.  In that 
particular case, the model is supported by the symmetric profile of the broad 
Pa$\alpha$ emission line, as well as the fact that the orientation of the radio 
jet is almost perfectly perpendicular to the gradient of the photocentres.  With 
more limited knowledge available for IRAS 09149$-$6206, it is not clear 
{\it a priori} whether the velocity gradient we observe reflects rotational or 
radial motion of the BLR.  As such, we adopt a flexible BLR model with a 
generalised prescription of the BLR dynamics \citep{Pancoast2014MNRASa}.  The 
specific code we use here was developed by \cite{Stock2018} and already adopted 
in the analysis of 3C~273 (GC18).

Following a general description of the model and its various parameters in 
Sec.~\ref{ssec:blrmod}, we compare the fits for two different implementations, 
allowing in both cases for an offset between the continuum and the BLR as 
discussed above.  For the first case, in Sec.~\ref{sec:rotmod}, we reduce the 
model to circular orbits as was applied in the case of 3C~273.  For the 
second case, in Sec.~\ref{sec:outmod}, we apply the full model, allowing for 
radial motions.  Sec.~\ref{sec:blrvamp} then checks the consistency of both 
models with the observed differential visibility amplitudes.  Finally, a 
comparison of the fits in Sec.~\ref{sec:compmod} shows that the goodness of fit 
does not indicate a preference based on the data alone, and it is instead the 
astrophysical implication that leads to a preference of one model.

\begin{table*}
\centering
\renewcommand{\arraystretch}{1.2}
\begin{tabular}{c | l | l}
\hline\hline
$R_{\rm BLR}$ & Mean radius of the BLR & $\mathrm{LogUniform(10^{-4}, 10\,pc)}$  \\
$F$ & Minimum radius of the BLR in units of $R_{\rm BLR}$ & $\mathrm{Uniform(0, 1)}$ \\
$\beta$ & Unit standard deviation of BLR radial profile & $\mathrm{Uniform(0, 2)}$   \\
$\theta_\mathrm{o}$ & Angular thickness measured from the mid-plane & $\mathrm{Uniform(0, \pi/2)}$ \\
$i$ & Inclination angle & $\mathrm{Uniform}(\cos{i (0, \pi/3)})$  \\
$PA$ & Position angle of the line of nodes on sky (east of north) & $\mathrm{Uniform(0, 2\pi)}$ \\
$\kappa$ & Anisotropy of the cloud emission & $\mathrm{Uniform(-0.5, 0.5)}$ \\
$\gamma$ & Clustering of the clouds at the edge of the disk & $\mathrm{Uniform(1, 5)}$ \\
$\xi$ & Mid-plane transparency & $\mathrm{Uniform(0, 1)}$ \\
\mbh & Black hole mass & $\mathrm{LogUniform}(10^5, 10^{10}\,M_\odot)$ \\
$f_\mathrm{ellip}$ & Fraction of clouds in bound elliptical orbits & $\mathrm{Uniform(0, 1)}$ \\
$f_\mathrm{flow}$ & Flag for specifying inflowing or outflowing orbits & $\mathrm{Uniform(0, 1)}$ \\
$\theta_\mathrm{e}$ & Angular location for radial orbit distribution & $\mathrm{Uniform(0, \pi/2)}$ \\
$\sigma_{\rho, \mathrm{circ}}$ & Radial standard deviation for circular orbit distribution & $\mathrm{LogUniform(0.001, 0.1)}$ \\
$\sigma_{\Theta, \mathrm{circ}}$ & Angular standard deviation for circular orbit distribution & $\mathrm{LogUniform(0.001, 1)}$ \\
$\sigma_{\rho, \mathrm{radial}}$ & Radial standard deviation for radial orbit distribution & $\mathrm{LogUniform(0.001, 0.1)}$ \\
$\sigma_{\Theta, \mathrm{radial}}$ & Angular standard deviation for radial orbit distribution & $\mathrm{LogUniform(0.001, 1)}$ \\
$\sigma_\mathrm{turb}$ & Normalized standard deviation of turbulent velocities & $\mathrm{LogUniform(0.001, 0.1)}$ \\ \hline 
$\lambda_\mathrm{emit}$ & Central wavelength of the emission line & $\mathrm{Norm(2.2896, 0.002\,\micron)}$ \\
$f_\mathrm{peak}$ & Peak flux of the normalized line profile & $\mathrm{Uniform(0.05, 0.065)}$ \\
$(x_\mathrm{o}, y_\mathrm{o})$ & Offset of the origin of the BLR & $\mathrm{Uniform(-1, 1\,\mathrm{mas})}$ \\
\hline\hline
\end{tabular}
\caption{Parameters of the BLR model with definitions, priors, and units where appropriate (all angles are in radians). 
The priors for most parameters are specified here in one of two ways. 
Uniform(min,max) denotes uniform sampling over the range specified. 
LogUniform(min,max) indicates that the logarithm of the parameter is sampled 
uniformly over the logarithm of the range. 
The prior for the inclination angle ($i$) is set between 0 and 
$\pi/2$ so that $\cos{i}$ is uniformly sampled between 0 and 1.  The prior for the central 
wavelength of the emission line ($\lambda_\mathrm{emit}$) follows a 
Gaussian distribution centered at 2.2896~\micron\ and with a standard deviation 
0.002~\micron.
}
\label{tab:prior}
\end{table*}

\subsection{The generalised BLR model}
\label{ssec:blrmod}

The generalised BLR model was developed by \citet[][P14]{Pancoast2014MNRASa}, 
with the original purpose to model the spectra and light curves from RM 
campaigns.  The P14 model describes the BLR as a large number of non-interacting 
clouds and includes a large number of parameters, summarised in 
Table~\ref{tab:prior}, that define the position and motion of each of those 
clouds.  Below, we briefly describe how these affect the geometry and dynamics 
of the model.

The first set of parameters defines the locations of the clouds.  Their 
distances from the black hole are given as
\begin{equation}
r = R_\mathrm{S} + F\, R_\mathrm{BLR} + g\, (1-F)\, \beta^2\, R_\mathrm{BLR},
\end{equation}
where $R_\mathrm{S}=2G\mbh/c^2$ is the Schwarzschild radius, $R_\mathrm{BLR}$ is 
the mean radius, $F=R_\mathrm{min}/R_\mathrm{BLR}$ is the fractional inner 
radius, $\beta$ is the shape parameter, and $g=p(x|1/\beta^2, 1)$ is drawn 
randomly from a Gamma distribution
\begin{equation}
p(x|a,b) = \frac{x^{a-1}e^{-x/b}}{\Gamma(a)\,b^a},
\end{equation}
where $\Gamma(a)$ is the gamma function.  Using a shape parameter in this way 
provides enough flexibility to reproduce several qualitatively different radial 
distributions, namely a Gaussian ($0<\beta<1$), exponential ($\beta=1$), or 
heavy-tailed ($1<\beta<2$) profile.  The angular distribution of the clouds is 
given by 
\begin{equation}\label{eq:theta}
\theta=\arccos{(\cos{\thetao} + (1-\cos{\thetao}) \times U^\gamma)}, 
\end{equation}
where $\thetao$ is the angular thickness of the distribution (defined as the 
angle between the mid-plane and the upper edge of the distribution) and $U$ is a 
random number drawn uniformly between 0 and 1.  Setting $\gamma>1$ concentrates 
more clouds closer to the maximum angular height $\thetao$.  The structure is 
viewed at an inclination angle $i$ (where $i=0^\circ$ is defined to be face-on) 
and rotated in the plane of the sky so that the line of nodes is at position 
angle $PA$ (measured east of north).  A weight is assigned to each cloud to 
represent the relative strength of its emission, and is defined as
\begin{equation}
w = 0.5 + \kappa \cos \phi,
\end{equation}
where $\kappa$ is a parameter in the range ($-0.5$, 0.5) reflecting any 
anisotropy of the emission, and $\phi$ is the angle between the line of sight 
from the cloud to the observer and to the central ionising source.  The 
mid-plane transparency is modelled using the parameter $\xi$ which controls the 
fraction of clouds located behind the equatorial plane.  If $\xi=1$ then the 
clouds are evenly distributed on either side of the equatorial plane, while 
$\xi=0$ means that all the clouds are in front of it.

The remainder of the parameters define the kinematics of the clouds, under the 
assumption that their motions are governed entirely by the gravitational 
potential of the black hole.  A fraction, $f_\mathrm{ellip}$, of the clouds are 
put on bound elliptical orbits.  The rest are placed on much more elongated 
orbits, which are dominated by radial motion. A single parameter, 
$f_\mathrm{flow}$, is used as a binary switch to control whether the radial 
motion is inflow ($0<f_\mathrm{flow}<0.5$) or outflow ($0.5<f_\mathrm{flow}<1$). 

For the bound elliptical orbits, radial and tangential velocities, 
$v_\mathrm{r}$ and $v_\phi$, are drawn randomly from a distribution centred on 
the point $\{0, \vcirc\}$ in the $v_\mathrm{r}$--$v_\phi$ plane (see Fig.~2 in 
\citealt{Pancoast2014MNRASa}), where $\vcirc=\sqrt{G\mbh/r}$ is the 
circular velocity.  The distribution itself follows an ellipse in the 
$v_\mathrm{r}$--$v_\phi$ plane that is defined as
\begin{equation}\label{eq:ellip}
\frac{v_\mathrm{r}^2}{2 v_\mathrm{circ}^2} + \frac{v_\phi^2}{v_\mathrm{circ}^2} = 1,
\end{equation}
and is defined to be a Gaussian with standard deviation 
$\sigma_{\Theta,\mathrm{circ}}$ along the ellipse and 
$\sigma_{\rho,\mathrm{circ}}$ perpendicular to it.

Because the ellipse naturally connects points around $\{0,\vcirc\}$ 
corresponding to circular orbits, with points around 
$\{\pm\sqrt{2}\vcirc,0\}$ corresponding to highly elongated orbits, the orbits 
dominated by radial motion can be defined in a similar way.  In this case, there 
is an additional parameter $\theta_\mathrm{e}=\arctan{(|v_\phi/v_\mathrm{r}|)}$ 
that defines the angular location around the ellipse where the distribution is 
centred.  If $\theta_\mathrm{e}$ is close to $\pi/2$, the orbit distribution is 
centred around $\{0,\vcirc\}$ exactly as for the bound elliptical orbits and 
there is very little inflow or outflow.  As $\theta_\mathrm{e}$ approaches 0, 
the centre of the distribution shifts to $\{\pm\sqrt{2}\vcirc,0\}$ where orbits 
are dominated by radial motion at the escape velocity.  The distribution is 
defined around the point on the ellipse corresponding to $\theta_\mathrm{e}$ as 
a Gaussian with standard deviation $\sigma_{\Theta,\mathrm{radial}}$ along the 
ellipse and $\sigma_{\rho,\mathrm{radial}}$ perpendicular to it.  The units of 
$\sigma_{\rho,\mathrm{circ}}$ and $\sigma_{\rho,\mathrm{radial}}$ are given in 
terms the circular velocity of the clouds, while 
$\sigma_{\Theta,\mathrm{circ}}$ and $\sigma_{\Theta,\mathrm{radial}}$ are given 
as a fraction of $\pi$.  The last velocity component is a random velocity 
$v_\mathrm{turb}$ describing the macroturbulence.  This is randomly sampled from 
a Gaussian distribution with standard deviation 
$\sigma_\mathrm{turb}v_{\rm circ}$, and added to the line-of-sight velocity of 
each cloud.

The full relativistic Doppler effect and gravitational redshift are also taken 
into account in generating the spectrum and differential phases.  For each 
cloud, the intrinsic line width is assumed to be negligible.  The 
wavelength of the line is shifted from $\lambda_\mathrm{emit}$ to 
$\lambda_\mathrm{obs}$ (both in the observed frame) by
%
\begin{equation}
\lambda_\mathrm{obs} = \lambda_\mathrm{emit} 
\frac{1 - \frac{v_\mathrm{los}}{c}}{\sqrt{1 - \frac{v^2}{c^2}}}
\frac{1}{\sqrt{1 - \frac{R_\mathrm{s}}{r}}},
\end{equation}
where $v_\mathrm{los}$ is the total line-of-sight velocity and $v$ is the total 
velocity.  Finally, we bin the clouds in the spectral channels according to 
their $\lambda_\mathrm{obs}$ and sum their weights to derive the normalised line 
profile.  The profile is then scaled by $f_\mathrm{peak}$, so that it has the 
same normalisation as the continuum.  The projected coordinates perpendicular to 
the line of sight are averaged in each bin according to the cloud weights to 
derive the photocentre of each spectral channel.  The differential phase is then 
calculated with Equation~(\ref{eq:dphi_fit_text}) with,
\begin{equation}
\phi_{\mathrm{BLR},\lambda} = -2\pi\, \frac{f_\lambda}{1 + f_\lambda} 
\vec{u} \cdot \left( \frac{\sum_i w_i \vec{x}_i}{\sum_i w_i} \right),
\end{equation}
where $w_i$ and $\vec{x}_i$ are the weight and coordinate of the $i$th cloud of 
the BLR with $\lambda_\mathrm{obs}$ within the wavelength channel $\lambda$.

Our prior assumptions on all these parameters, as given in 
Table~\ref{tab:prior}, generally follow the choices of 
\cite{Pancoast2014MNRASa}.  The main exception is the inclination angle, which 
we require to be below 50\degree, because the BLR of a type~1 AGN is expected to 
be relatively face-on.\footnote{Allowing $i$ to vary over the full range of 
0\degree--90\degree\ leads to multiple modes in the posterior distribution for 
the outflow model described in Sec.~\ref{sec:outmod} and a preference for a 
highly inclined geometry inconsistent with the expectation for a Seyfert~1.
We therefore apply a more restrictive prior to the inclination angle. Our 
results are unaffected if we increase the upper limit of $i$ to, for example, 
60\degree.  We choose $i<50\degree$ because the posterior probability density 
starts to rise towards higher values of $i$.}  For the prior on 
$\lambda_\mathrm{emit}$, we adopt a Gaussian distribution centred at 
2.2896~\micron\ based on the redshift measured by \cite{Perez1989AA}. The 
standard deviation of the Gaussian distribution is 0.002~\micron, which is the 
width of the spectral channel and the equivalent $\sigma$ of the line spread 
function for the GRAVITY science spectrograph.  Three additional parameters that 
we include are the peak flux $f_{\rm peak}$ of the emission line normalised to 
the continuum, and the offset $\{x_{\rm o},y_{\rm o}\}$ for the origin of the 
BLR.  The prior range adopted for $f_\mathrm{peak}$ is 0.05--0.065.  The offset 
of the BLR, $\{x_\mathrm{o}, y_\mathrm{o}\}$, is allowed to vary by up to $-1$ 
to 1~mas in both directions.

\begin{figure*}
\centering
\includegraphics[height=0.2\textheight]{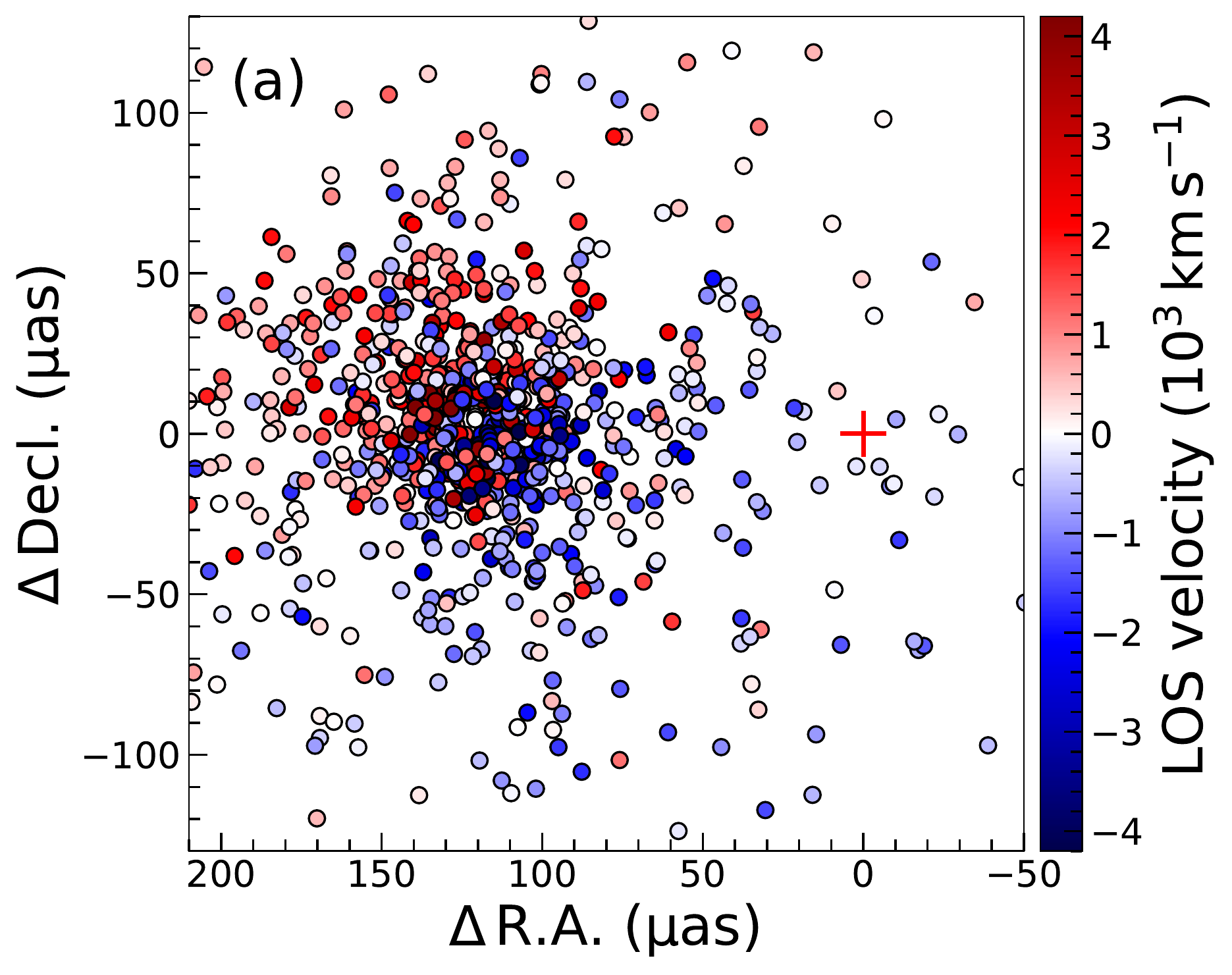}
\includegraphics[height=0.2\textheight]{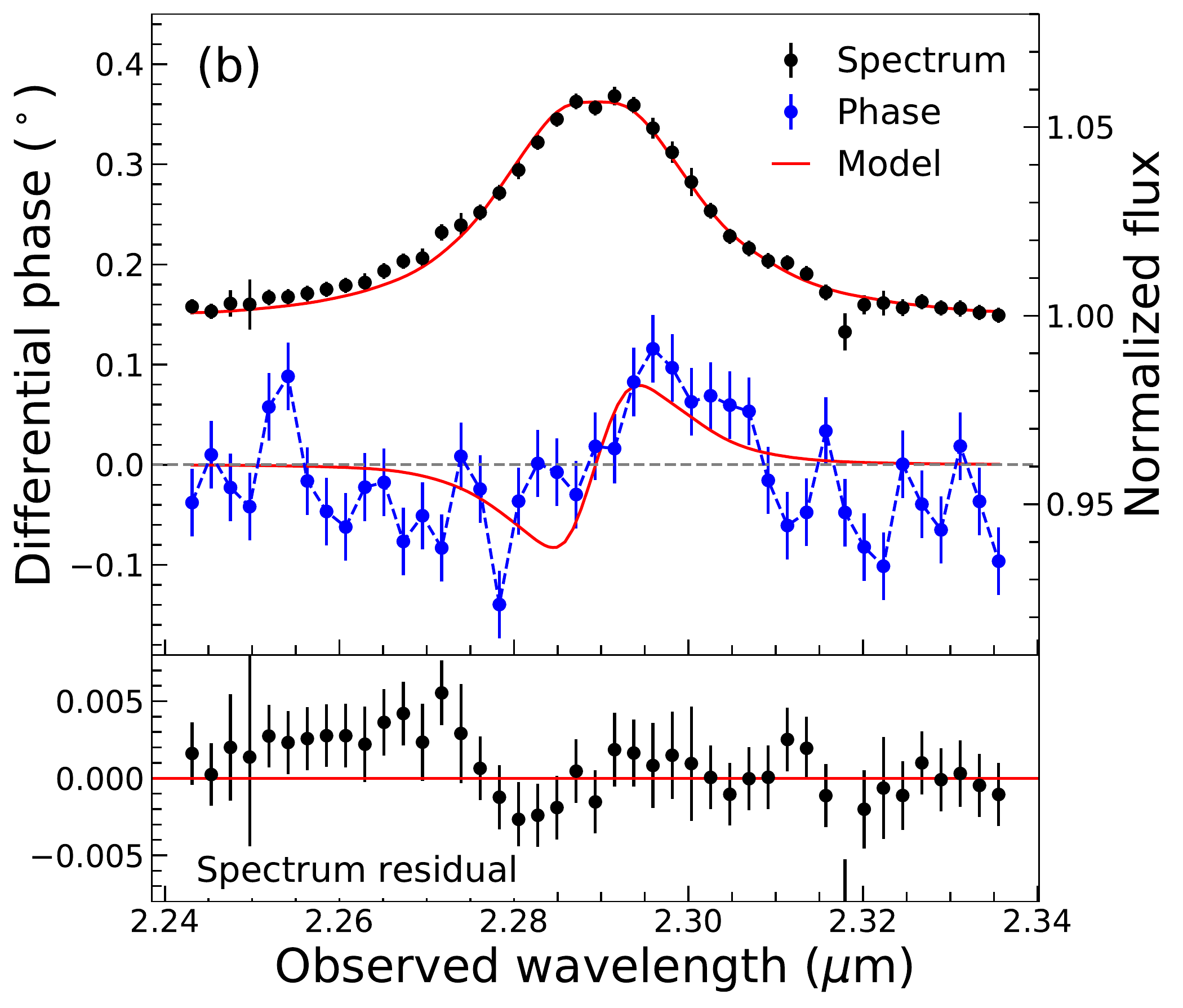}
\includegraphics[height=0.2\textheight]{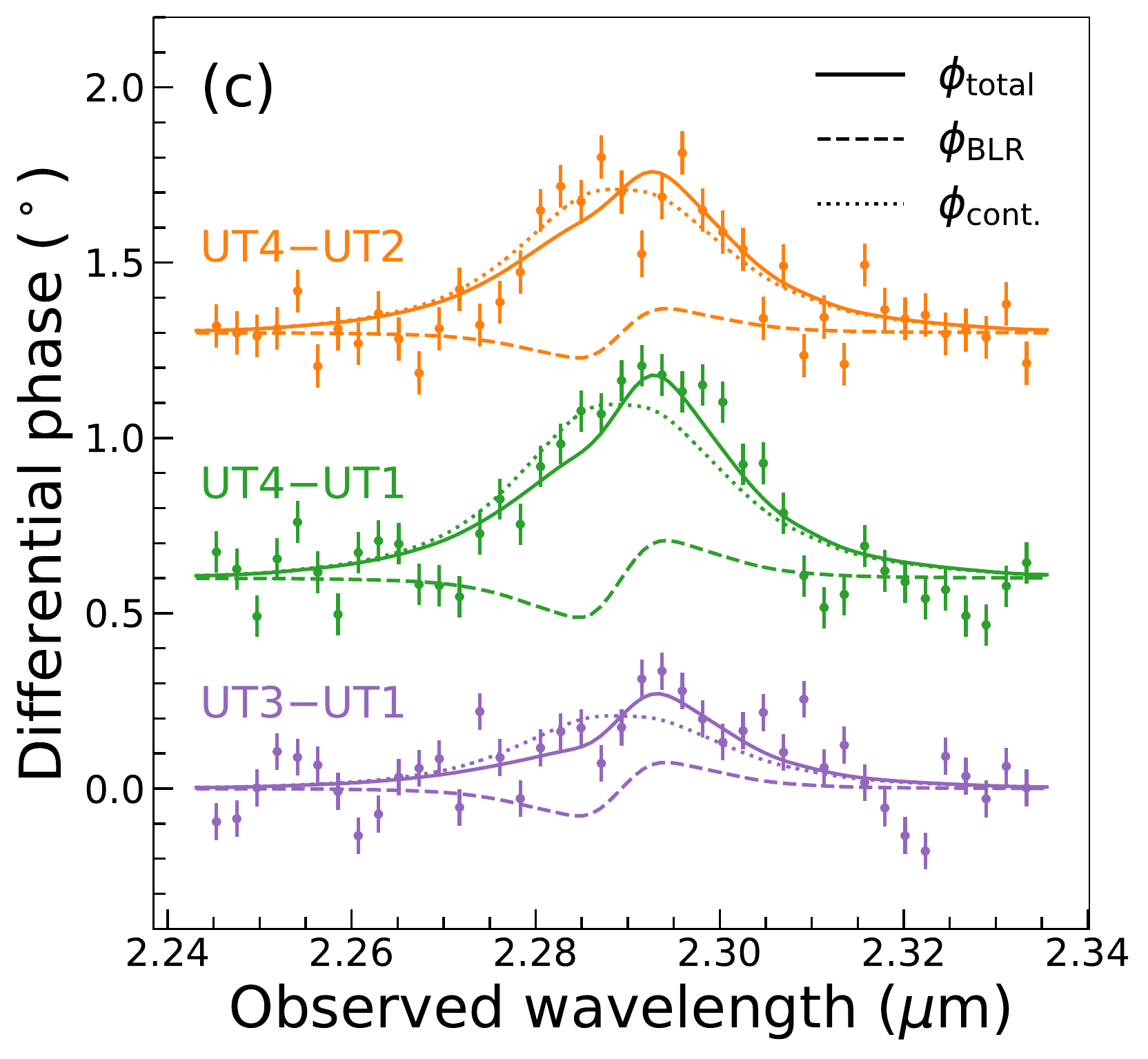}
\caption{(a) The cloud distribution of the best-fit Keplerian 
model.  Each circle represents one cloud, colour-coded by the line-of-sight 
velocity.  The zero velocity is defined at the best-fit $\lambda_\mathrm{emit}$. 
The red plus sign at the origin represents the photocentre of the 
continuum.  The orientation of the model is consistent with the observed 
photocentre gradient.  (b) The observed averaged differential phase from 
UT4$-$UT2, UT4$-$UT1, and UT3$-$UT1 after removing the `continuum phase' signal 
(blue) compared to the averaged differential phase from the best-fit BLR model 
(red).  These baselines were chosen since they contain the strongest `S-shape' 
signal. Above, the observed line profile (black) is compared with the model line 
profile (red).  The residual of the spectrum subtracting the model line 
profile is displayed in the lower panel.  The left $y$ axis corresponds to the 
averaged differential phase, while the right $y$ axis corresponds to the line 
profile.  (c) The differential phase data and the best-fit models (solid lines) 
of the three baselines that show the strongest signal of the BLR component 
(dashed lines).  The phase in panel (b) is calculated by averaging the phases of 
these three baselines after subtracting the best-fit continuum phases (dotted 
lines). 
}
\label{fig:blr_kp}
\end{figure*}

\subsection{Model with Circular Keplerian Rotation}
\label{sec:rotmod}

The P14 model described above can easily be reduced to circular Keplerian rotation. 
Setting $\kappa=0$ and $\xi=1$ ensures that the clouds are equally weighted 
(i.e. emitting isotropically) and are distributed uniformly above and below the 
equatorial plane.  Instead of using Equation~(\ref{eq:theta}) to determine 
the initial angular distribution of the clouds, the circular Keplerian 
model distributes them uniformly between 0 and $\thetao$ (GC18).  To produce 
bound circular orbits, we set $f_\mathrm{ellip}=1$, 
$\sigma_{\rho,\mathrm{circ}}=0$, and $\sigma_{\Theta,\mathrm{circ}}=0$.  
Finally, setting $\sigma_\mathrm{turb}=0$ ensures that there is no additional 
turbulence.  We refer to this model as the ``Keplerian model'', hereafter, 
for simplicity.

The flux and differential phase spectra of IRAS~09149$-$6206 are fit reasonably 
well by the Keplerian model, as shown in Fig.~\ref{fig:phs} for the averaged 
bins (see also Fig.~\ref{fig:phs_kp} for the individual bins).  The most 
prominent phase signals in the UT4$-$UT2, UT4$-$UT1, and UT3$-$UT1 baselines are 
primarily due to the continuum phase produced by the offset between the BLR and 
the centre of the continuum emission.  For this model, the BLR is 
$\sim 120\,\mu$as east of the continuum centre.  The cloud distribution of the 
model is shown in Fig.~\ref{fig:blr_kp}\,(a), and the corresponding photocentres 
in Fig.~\ref{fig:phc}\,(b) are qualitatively consistent with those reconstructed 
from the data.  The best fitting parameters summarised in Table~\ref{tab:blr} 
indicate that a rather face-on disk with $i \approx 21\degree$ is favoured by 
the data, which is consistent with the inclination found from an upcoming dust 
reverberation study (S. H\"{o}nig et al. in preparation).  A low inclination is 
consistent with the Seyfert~1 classification and, as one would expect, low 
inclinations are generally inferred when fitting RM data of other objects.  The 
disk is also very thick, with $\thetao \approx 71\degree$.  \cite{Grier2017ApJb} 
have suggested that, when fitting RM data, the thickness $\thetao$ is always 
similar to the inclination angle $i$, due to degeneracy between the two 
quantities. Interferometry breaks this degeneracy.  Our derived values for these 
parameters are significantly different, and the posterior distributions in 
Fig.~\ref{fig:cnr_kp} show no particular coupling.  The mean radius of 
$R_\mathrm{BLR}=65\,\mu$as corresponds to 0.075\,pc.  The posterior 
distributions in Fig.~\ref{fig:cnr_kp} indicate that there is some degeneracy 
between BLR radius, black hole mass, and the inclination angle, which is 
consistent with previous studies (\citealt{Rakshit2015MNRAS}; GC18).  Although 
the inferred line centre $\lambda_\mathrm{emit}=2.2892\,\micron$ corresponds to 
a velocity offset of $\Delta v_\mathrm{BLR}=-56\,\mathrm{km\,s^{-1}}$ with 
respect to the systemic velocity from the \OIII\ rotation curve 
\citep{Perez1989AA}, the uncertainty of $\Delta v_\mathrm{BLR}$ is large enough 
that the modelled line centre is fully consistent with it.

In order to display the phase signal specific to the BLR, we subtract 
the best-fit continuum phase from the data in the three longest baselines 
(UT4$-$UT2, UT4$-$UT1, and UT3$-$UT1) as shown in Fig.~\ref{fig:blr_kp}\,(c), 
and then average them.  The residual BLR signal, shown in 
Fig.~\ref{fig:blr_kp}\,(b), exhibits the expected `S-shape' profile for a 
rotating structure.  Based on the analysis in Appendix~\ref{apd:phase}, and 
taking into account that several baselines were combined, the uncertainty of 
this phase is expected to be below 0.03\degree\ per spectral channel.  As such, 
even though the $\sim 0.1 \degree$ BLR signal is several times weaker than the 
continuum phase signal, it remains a significant detection.  We 
note that the `S-shape' profile is due to specifically fitting with the 
Keplerian model. The significance of the BLR phase signal is 
constrained model-independently with the reconstructed photocentres 
(Sec.~\ref{sec:blrpos}).  Finally, we also need to consider the fit to the 
spectral line profile.  The Keplerian model is only able to generate a symmetric 
line profile, which also means that $\lambda_\mathrm{emit}$ is close to the 
wavelength of the peak of the line profile.  However, the observed line profile 
of IRAS~09149$-$6206 is slightly asymmetric.  As such, although the line profile 
is reasonably well matched by the model, a difference between the model and 
data, especially in the blue wing (e.g., 2.24--2.29~\micron), is 
apparent in Fig.~\ref{fig:blr_kp}\,(b).  This issue is addressed in the next 
Section.

\begin{figure*}
\centering
\includegraphics[height=0.2\textheight]{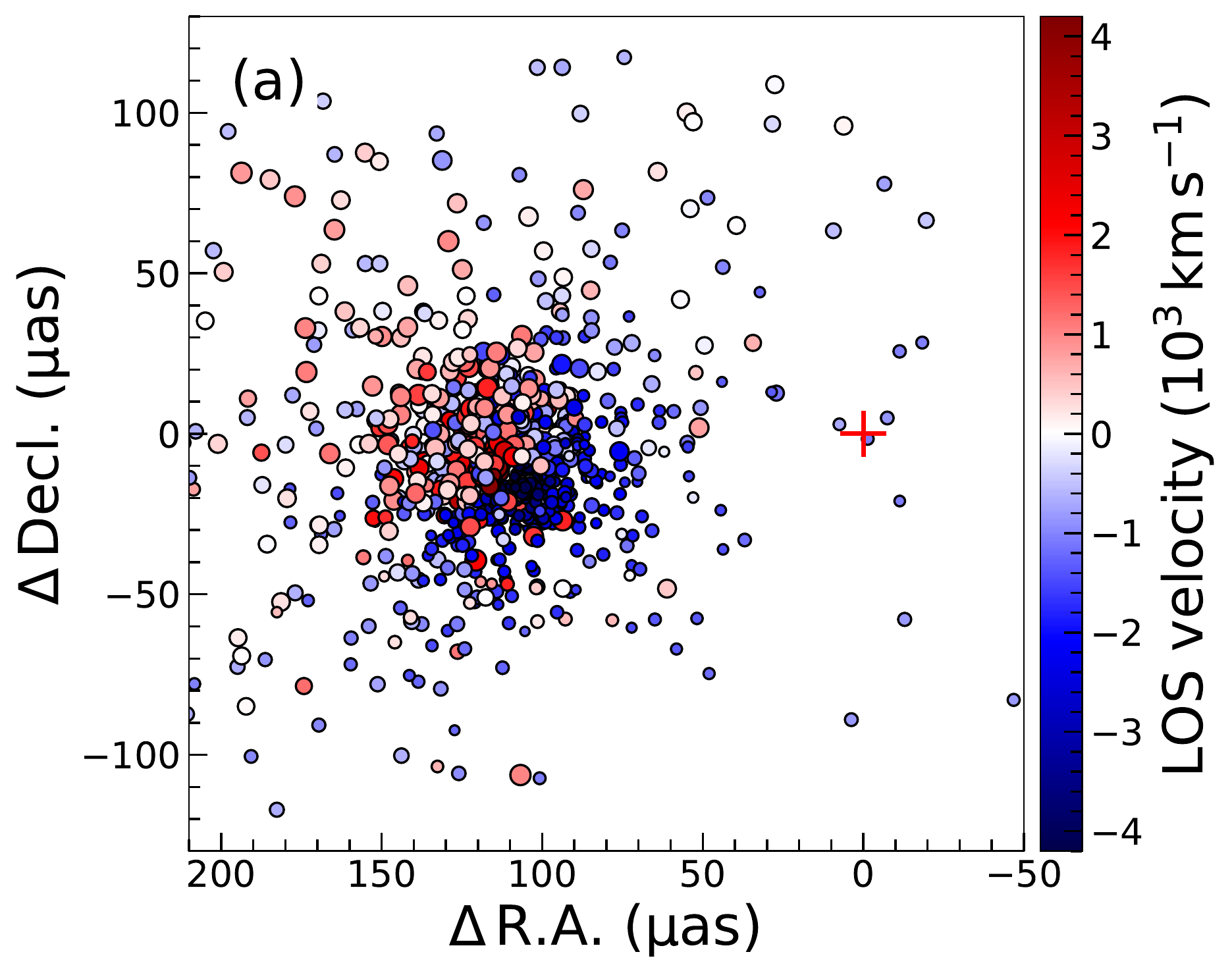}
\includegraphics[height=0.2\textheight]{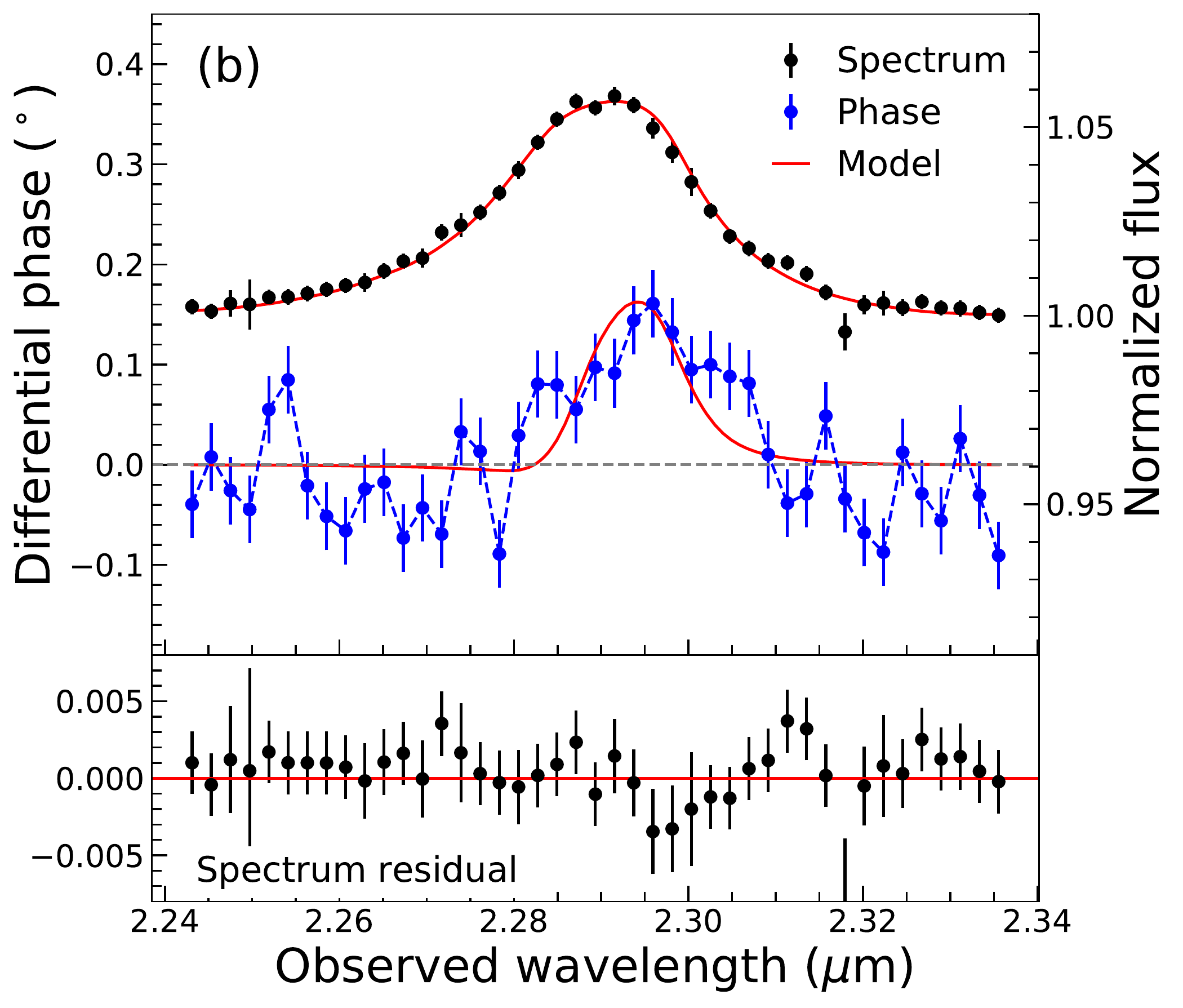}
\includegraphics[height=0.2\textheight]{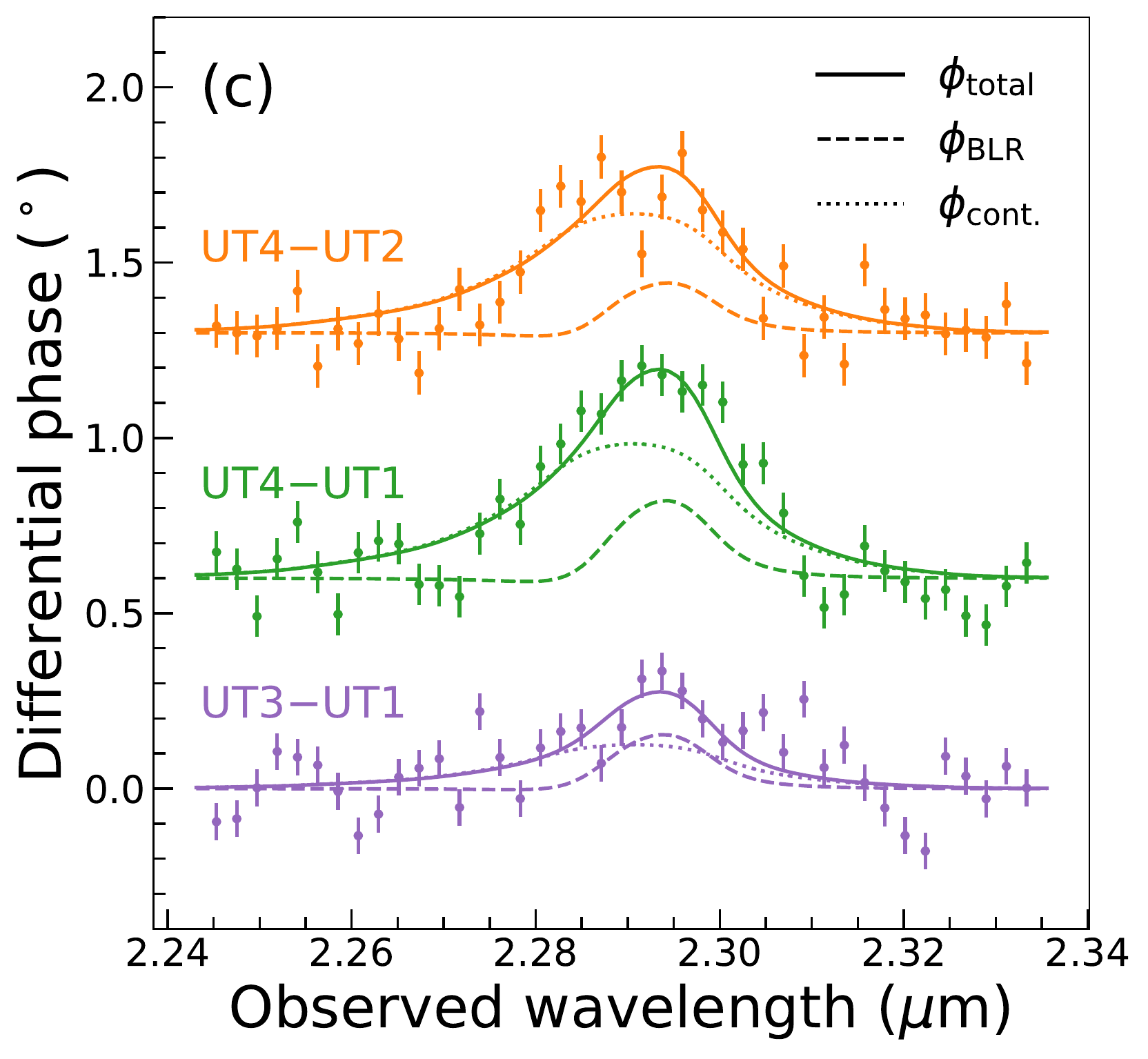}
\caption{(a) The cloud distribution of the best-fit outflow 
model. The symbols and lines are the same as for Fig.~\ref{fig:blr_kp}, except that in panel (a) the sizes of the circles scale with the weight of the cloud.}
\label{fig:blr_of}
\end{figure*}

\begin{figure*}
\centering
\includegraphics[height=0.25\textheight]{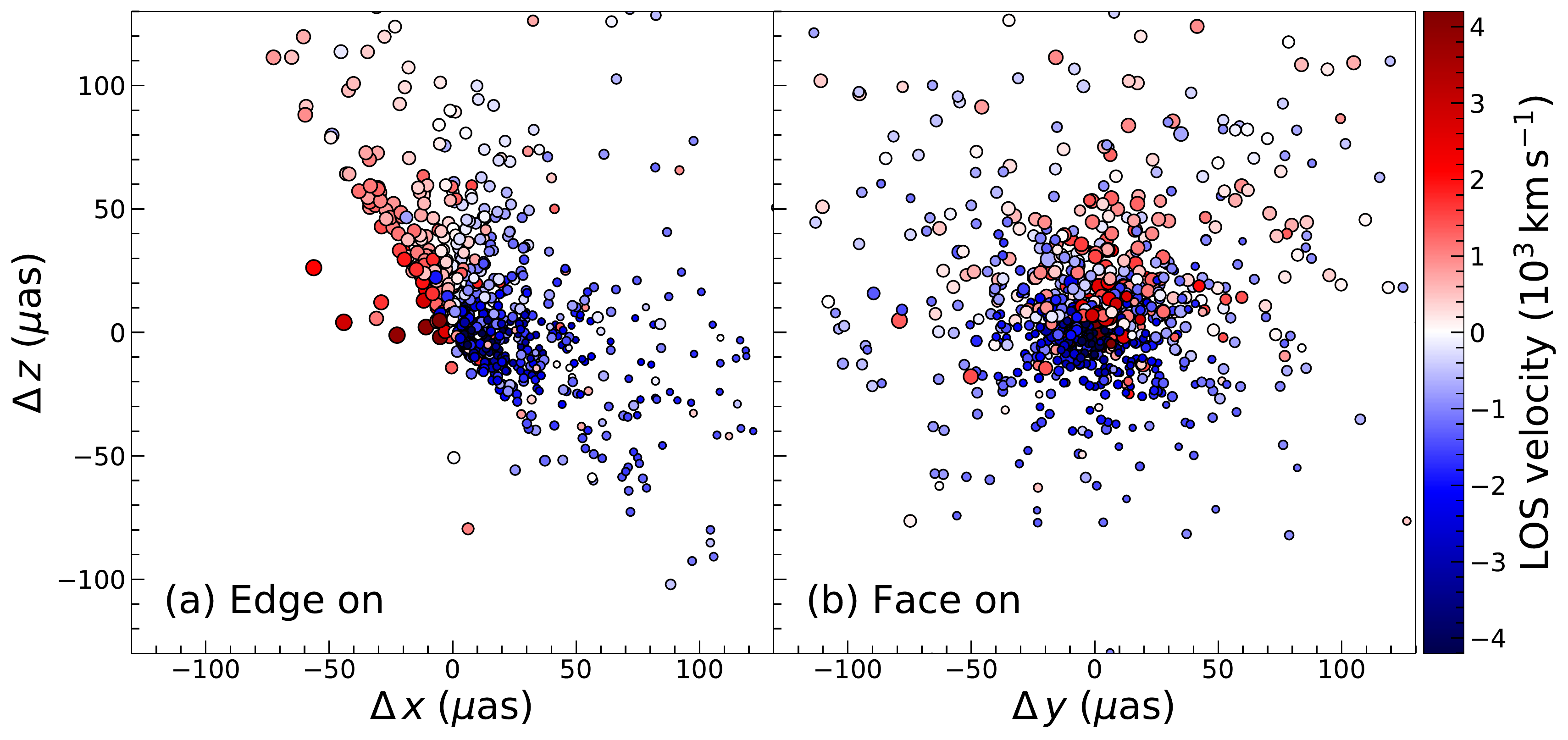}
\caption{The (a) edge-on and (b) face-on view of the best-fit outflow
model, except that the PA is adjusted to 180\degree\ and the BLR is moved back to 
the origin of the plot for clarity.  
In both panels, the colour coding represents the line-of-sight velocity for each cloud.
Following 
\cite{Pancoast2014MNRASa,Pancoast2014MNRASb}, $\Delta x$ is along the line of 
sight, $\Delta y$ is the direction of the right ascension, and $\Delta z$ is the 
direction of declination.}
\label{fig:blr_view}
\end{figure*}

\subsection{Model including Radial Motion}
\label{sec:outmod}

\begin{table*}
\centering
\renewcommand{\arraystretch}{1.5}
\begin{tabular}{l | c | c }
\hline\hline
                          Parameters &  Circular Keplerian model          &           Outflow model \\ \hline
$R_\mathrm{BLR}\,(\mu\mathrm{as})$   & $65_{-39}^{+30}$                   & $50_{-11}^{+38}$        \\
$R_\mathrm{min}\,(\mu\mathrm{as})$   & $8.3_{-7.4}^{+13.1}$               & $6.8_{-5.0}^{+7.5}$     \\
$\beta$                              & $1.17_{-0.33}^{+0.30}$             & $1.27_{-0.32}^{+0.17}$  \\
$\theta_\mathrm{o}\,(^\circ)$        & $71_{-27}^{+16}$                   & $61_{-20}^{+20}$        \\
$i\,(^\circ)$                        & $21_{-8}^{+20}$                    & $35_{-10}^{+13}$         \\
PA ($^\circ$ E of N)                 & $130_{-34}^{+29}$                  & $219_{-37}^{+27}$       \\
$\kappa$                             & \nodata                            & $-0.32_{-0.17}^{+0.44}$ \\
$\gamma$                             & \nodata                            & $1.27_{-0.20}^{+2.40}$  \\
$\xi$                                & \nodata                            & $0.05_{-0.04}^{+0.30}$  \\
Offset ($\mu$as)                     & $\left(121.6_{-9.7}^{+6.5}, 4.5_{-8.6}^{+8.4}\right)$ 
                                     & $\left(109.2_{-21.8}^{+11.5}, -13.9_{-17.4}^{+12.3}\right)$ \\ \hline
$\log\,(\mbh/M_\odot)$               & $8.06_{-0.57}^{+0.41}$             & $7.70_{-0.18}^{+0.41}$   \\
$f_\mathrm{ellip}$                   & 1                                  & $0.19_{-0.17}^{+0.35}$   \\
$f_\mathrm{flow}$                    & \nodata                            & $0.71_{-0.20}^{+0.26}$   \\
$\theta_\mathrm{e}$                  & \nodata                            & $5.0_{-4.3}^{+22.9}$    \\
$\log\,\sigma_\mathrm{turb}$         & 0                                  & $-1.87_{-1.04}^{+0.80}$  \\ 
$\Delta v_\mathrm{BLR}\,(\mathrm{km\,s^{-1}})$  
                                     & $-56_{-67}^{+78}$                  & $380_{-356}^{+208}$      \\ \hline
$\chi_\mathrm{r}^2$                  & 1.39                               & 1.38                     \\
$\ln\,K$                             & 0                                  & $7.1 \pm  0.2$           \\
$\Delta$ AIC                         & 0                                  & -12.6                    \\
$\Delta$ BIC                         & 0                                  & 44.0                     \\
\hline\hline
\end{tabular}
\caption{The inferred maximum {\it a posteriori} value and central 95\% credible interval 
for the modelling of the spectrum and differential phase of IRAS~09149$-$6206.  
$\Delta v_\mathrm{BLR}$ is the difference between the velocity derived from the best-fit $\lambda_\mathrm{emit}$ and the systemic velocity 
based on the \OIII\ line \citep{Perez1989AA}.
$\chi_\mathrm{r}^2$ is the reduced $\chi^2$ of the models with the best-fit 
parameters.  The Bayes' factor, AIC, and BIC are relative to the Keplerian model.
}
\label{tab:blr}
\end{table*}

To be able to fit the asymmetries in the emission line profile, we apply the 
full P14 model.  The best fitting parameters of the P14 model 
(Table~\ref{tab:blr}) include $f_\mathrm{ellip} \approx 0.2$ indicating 
that the majority of the clouds are on orbits with a dominant radial component, 
$\theta_{\rm e} \approx 5\degree$ indicating that the orbits are sufficiently 
elongated that the radial motion is very close to the escape velocity, and 
$f_\mathrm{flow}>0.5$ indicating that this radial motion is outward.  Together, 
these indicate that, although it is not required {\it a priori}, the 
configuration of the model preferred by the data is very much dominated by 
outflow. As such, hereafter we call this the `outflow model'.

As before, the phase signals shown in Fig.~\ref{fig:phs} (see also 
Fig.~\ref{fig:phs_of} for the individual bins) are dominated by the continuum 
phase.  The BLR offset of $\sim 110\,\mu$as, which can be seen in 
Fig.~\ref{fig:blr_of}\,(a), is statistically consistent with that of the 
Keplerian model.  The modest difference is due to the different BLR phase 
signals (Fig.~\ref{fig:blr_of}(b)).  The orientation and gradient of the 
photocentres in Fig.~\ref{fig:phc}\,(c), are also consistent with the data.  
The outflow model indeed better fits the line profile compared to the Keplerian 
model.  No systematic residual is seen in Fig.~\ref{fig:blr_of}(b), especially in the blue wing.
Because the two models fit the differential phase data equally well, as shown in 
Fig.~\ref{fig:phs}, the total goodness of fit for the two models are nearly indistinguishable (Sec.~\ref{sec:compmod}).

The mean radius of $R_\mathrm{BLR}=50\,\mu$as is slightly smaller than, but 
statistically consistent with, that of the Keplerian model.  The cloud radial 
distribution given by $\beta=1.27$ prefers to be exponential or heavy-tailed, 
with a small inner radius of $R_\mathrm{min}=6.8\,\mu$as.  The model has 
$\mathrm{PA} \approx 219\degree$, which is 90\degree\ different from Keplerian 
model.  The reason is simply that the BLR kinematics, and hence the orientation 
of the velocity gradient, are now dominated by radial motion rather than 
rotation.  This model also prefers anisotropic emission from the clouds, with 
$\kappa=-0.32$ indicating that line emission is stronger from the inner 
illuminated side of the clouds and hence the far side of the distribution, 
similar to many of the results inferred from RM data.  The parameter 
$\sigma_\mathrm{turb}=0.013$ implies that additional macroscopic turbulence is 
not significant.  As for the Keplerian model, $\theta_0 \sim 61$\degree\ 
indicates that the cloud trajectories are distributed over a wide range of 
angles from the mid-plane.  We also note, similarly to the Keplerian model, that 
the thickness and inclination have somewhat different values, and there is no 
evidence for degeneracy between them in the posterior distributions shown in 
Fig.~\ref{fig:cnr_of}.  Indeed, except for the greater thickness of the BLR in 
IRAS~09149$-$6206, the configuration is rather similar to that inferred from RM 
data for Arp~151 \citep{Pancoast2014MNRASb}, Zw~229$-$015 
\citep{Williams2018ApJ}, or Mrk~142 \citep{Li2018ApJ}.  Finally, the best fit 
line centre is $\lambda_\mathrm{emit}=2.2923\,\micron$ which corresponds to an 
offset $\Delta v_\mathrm{BLR}=380\,\mathrm{km\,s^{-1}}$ from the systemic 
velocity.  A discussion of whether this is physically plausible is deferred to 
Sec.~\ref{sec:compmod}. 

One of the most important aspects of this model is that the differential phase 
of the BLR component is very different to the `S-shape' seen in the Keplerian 
model.  It is clear from Fig.~\ref{fig:blr_of}\,(b) that the continuum 
subtracted phase data fitted by the BLR model is dominated by a positive signal 
on the red-shifted side of the line profile.  Fig.~\ref{fig:blr_of}\,(c) 
illustrates the decomposition of the BLR phase and continuum phase components.  
The asymmetric BLR phase signal is produced by two main effects: (i) The BLR 
kinematics are dominated by outflow (as discussed above) and (ii) 
$\xi \approx 0$ means that the mid-plane is opaque.  The impact of this second 
effect is discussed below in the context of the distribution and motions of the 
clouds.

The edge-on and face-on views of the cloud distribution presented in  
Fig.~\ref{fig:blr_view} can shed more light on the role of the mid-plane 
obscuration in generating the positive phase signal.  The edge-on view clearly 
shows that the cloud distribution extends far above the mid-plane because 
$\thetao \approx 60\degree$; while $\xi \approx 0$ means that the mid-plane 
obscuration is so strong that there are few clouds below it.  As discussed 
above, our model is significantly dominated by outflowing clouds.  The 
blue-shifted clouds are on the near side towards the observer, while the 
red-shifted clouds are on the far side.  In addition, the anisotropy parameter 
$\kappa \approx -0.3$ means that the weighting applied to clouds on the far side 
is much larger than for the near side.  This effect is indicated in the figure 
by the size of the circles representing the clouds: on the side nearer the 
observer, the circles are much smaller than those on the far side.  The 
mid-plane obscuration and inclination angle together mean that, as is apparent 
in Fig.~\ref{fig:blr_of}\,(a), the blue-shifted clouds are distributed fairly 
symmetrically around the centre of the BLR.  This means that their photocentre 
is close to the BLR centre and the corresponding differential phases are close 
to 0\degree.  In contrast, the red-shifted clouds are primarily located to the 
northeast of the BLR centre, so the corresponding red-shifted channels show 
significant differential phase signal.  As a result, as shown in 
Fig.~\ref{fig:phc}, the origin of the BLR coincides with the blue-shifted 
channels instead of with the channel associated with the line peak.

\subsection{Model prediction of the differential visibility amplitude} 
\label{sec:blrvamp}

The measured differential visibility amplitude is useful to provide an 
independent check of the BLR model fits.  The differential visibility amplitude 
($V_{\rm diff}$) is derived by normalising the total visibility amplitude with 
the visibility amplitude of the continuum $V_c$.  In each spectral channel 
$\lambda$, this is
\begin{equation}
V_\mathrm{diff}(\lambda) = 
\frac{1 + f_\lambda V_\mathrm{BLR}(\lambda)/V_c(\lambda)}{1 + f_\lambda},
\end{equation}
where $V_\mathrm{BLR}$ is the visibility amplitude of the line emission of the 
BLR.  At the wavelength of the line emission, one will find $V_\mathrm{diff}>1$ 
if $V_\mathrm{BLR}/V_c>1$, this is if the BLR emission is more compact than the 
continuum emission.  This can be seen in Fig.~\ref{fig:vamp}, in particular for 
the UT4$-$UT1 baseline.  Similar results have been reported for 3C~273 
(GC18) and PDS~456 (GC20a).

The visibility amplitude of the continuum emission of IRAS~09149$-$6206 has 
already been studied in GC20a.  Using the visibility amplitude from 
the fringe tracker channel and the differential visibility from the science 
channel, we derived consistent FWHM sizes $0.54\pm0.05$~mas and 
$0.64\pm0.06$~mas, respectively, for a circular Gaussian profile.  This 
indicates that the continuum emission is only marginally resolved.  For 
consistency, we adopt a Gaussian profile with $\mathrm{FWHM}=0.6\,\mathrm{mas}$ 
to calculate
\begin{equation}
V_c(\lambda) = \exp{\left(-\frac{(\pi\, \mathrm{FWHM})^2 
(u^2 + v^2)}{4 \ln 2} \right)}.
\end{equation}
Following \cite{Waisberg2017ApJ}, we calculate $V_\mathrm{BLR}$ from the 
Keplerian and outflow models from the second moment of the source emission,
\begin{equation}
V_\mathrm{BLR}(\lambda) \approx 1 - \frac{2 \pi^2}{\mu_{00,\lambda}} 
\left( u^2 \tilde{\mu}_{20,\lambda} + v^2 \tilde{\mu}_{02,\lambda} + 
2uv \tilde{\mu}_{11,\lambda} \right),
\end{equation}
where $\mu_{00,\lambda}=\sum_\lambda\, w_i$ is the total intensity of the BLR 
line emission (zero-order moment), and summing up the weight ($w_i$ for the 
$i$th cloud) of all of the clouds that belong to each spectral channel 
$\lambda$.  In addition 
$\tilde{\mu}_{pq,\lambda} = \sum_\lambda\, w_i 
(l_i - l_{c,\lambda})^p (m_i - m_{c,\lambda})^q$ 
is the relative moment of the line emission with respect to the photocentre of 
the spectral channel
\begin{equation}
(l_{c,\lambda}, m_{c,\lambda}) = \left( \frac{\sum_\lambda w_i l_i}{\mu_{00}},
\frac{\sum_\lambda w_i m_i}{\mu_{00}} \right).
\end{equation}
The marginally resolved approximation here is the same one used in calculating 
the BLR differential phase. It is valid when the BLR  is not more than 
marginally resolved ($2\pi\,\vec{u}\cdot \vec{x} \ll 1$ or 
$R_{\rm BLR} \ll 1$~mas). We calculated the differential visibility amplitude 
for both models, and plot them over the data in Fig.~\ref{fig:vamp} taking into 
account the instrumental line spread function.  The predicted differential 
visibility amplitudes for the two models are almost indistinguishable, because 
the derived BLR size $R_{\rm BLR}$ is almost the same at $\sim 60\,\mu$as in 
both cases.  The clear signal in the UT4$-$UT1 baseline is consistent with the 
compact BLR size inferred from the differential phase data.  We note that 
although this is sensitive to the adopted size of the continuum emission, it 
produces qualitatively similar results for continuum FWHM in the range 
0.54--0.64~mas.  Additional details are discussed in Appendix~\ref{apd:test}.

\subsection{Comparing the models}
\label{sec:compmod}

In the preceding sections, we have discussed the Keplerian and outflow models 
individually from a phenomenological perspective.  The offset and the 
orientation of the velocity gradient of the models are qualitatively consistent 
with those derived directly from the observed data of IRAS~09149$-$6206. We have 
also shown that the size of the BLR in both cases is consistent with the 
differential visibility amplitudes, which provide a direct comparison to the 
measured size of the continuum.  Here we try to compare them, both in terms of a 
statistical perspective and with reference to the literature.  

We calculate the reduced $\chi^2$ ($\chi_\mathrm{r}^2$), Bayes factor ($K$), 
Akaike information criterion (AIC) and Bayesian information criterion (BIC), in 
order to compare the goodness of fits for the two models.  These quantities 
are described in more detail in Appendix~\ref{apd:bayes}, and their values are 
given in Table~\ref{tab:blr} (in terms of the difference of the outflow model 
with respective to the Keplerian model) together with the relevant set of best 
fitting parameters.  The $\chi_\mathrm{r}^2$ is almost exactly the same for both 
models.  In contrast, the other quantities do indicate a preference. A high 
Bayes factor and negative $\Delta$AIC favours the outflow model.  However, a 
negative $\Delta$BIC favours the Keplerian model. This difference reflects the 
different approaches that the criteria adopt, in penalising the free parameters 
versus the prior information. Using mock data generated from the inferred 
parameters with the Keplerian and outflow models (see Appendix~\ref{apd:test}), 
we confirm that the Bayes factor seems to provide a more reliable model 
selection for this case.

Unfortunately, the literature provides no useful information about the spatially 
resolved radio emission nor the gas kinematics that might be able to shed light 
on the interpretation of the BLR kinematics.  In addition, the early 
interferometric measurements 
\citep{Kishimoto2011AA,Burtscher2013AA,LopezGonzaga2016AA} barely resolve the 
mid-infrared continuum emission.

Nevertheless, we can compare the models from the point of view of the physical 
interpretation.  The key parameter that generates the asymmetric line profile 
for the outflow model is the mid-plane transparency, with $\xi \sim 0$ 
indicating that it is opaque.  In contrast, the mid-plane of the Keplerian model 
is, by our definition, fully transparent.  It is not immediately clear what 
might physically cause an opaque mid-plane.  Absorption by gas is not possible 
because any given cloud could only absorb a narrow range of velocities; and so 
it would have to be due to extinction by dust. But to obscure Br$\gamma$ 
emission requires an optical depth $\tau \sim 1$ at 2.2\,$\mu$m.  For a standard 
gas-to-dust ratio, this would require a column of 
$N_{\rm H} > 10^{22}$\,cm$^{-2}$ of dusty gas that is nominally located inside 
the dust sublimation radius. This might be possible for the BLR concept proposed 
by \cite{baskin18} in which, because the emission from the accretion disk is 
anisotropic, large ($\gtrsim0.3$\,$\mu$m) graphite grains can survive close to 
the disk plane at radial scales associated with the BLR.  Indeed, the model 
requires this, since it purports that the BLR may be a failed dusty wind --- 
failed because while dust opacity allows a wind to be launched, the dust 
sublimates once clouds move upwards.  In the context here, the key point is that 
it does imply a possible physical source for the mid-plane opacity without 
affecting the optical/UV emitting inner accretion disk.  However, this model 
would also imply that much of the hot dust should exist on the same spatial 
scales as the BLR, while our interferometric data (especially 
Fig.~\ref{fig:vamp}), indicate that the hot dust is much more extended.  As 
such, we would argue that it is difficult for this model to explain such a high 
mid-plane opacity in a way that is consistent with the data.

An alternative possibility arises because, for a BLR dominated by clouds on 
radial outflowing trajectories, it is not entirely clear whether the parameters, 
such as the inclination angle and the disk thickness, should still retain their 
original physical interpretation.  The best-fit outflow model tends rather to 
mimic a polar outflow, although intrinsically limited by its disk-like 
construct.  However, a truly polar outflow would be inconsistent with the broad 
concept described by \cite{elvis00} in which, because the outflow originates in 
a disk, it has a major rotational component (as observed in PG\,1700$+$518 by 
\citealt{young07}) and is directed at an angle significantly offset from the 
polar direction. Consistency with physically motivated rotating wind models 
\citep{everett05,keating12,Mangham2017MNRAS} is difficult to achieve for the 
outflow model here, which is strongly polar. On the other hand, the Keplerian 
model is plausibly consistent with a disk wind, because of the wide angle 
$\thetao$ above and below the mid-plane over which its cloud trajectories are 
distributed.

Another impact of the opacity of the mid-plane in the outflow model is to reduce 
the number of red-shifted clouds, which means that the resulting line emission 
is dominated by the blue-shifted clouds.  The model, therefore, prefers a large 
$\lambda_\mathrm{emit} \approx 2.2923\,\micron$.  This moves the line profile 
from the model back so that it matches the wavelength of the observed line 
profile.  The best fitting $\lambda_\mathrm{emit}$ corresponds to a 
$\sim 380\,\mathrm{km\,s^{-1}}$ shift with respect to the systemic velocity, 
which is slightly below the 2-$\sigma$ lower boundary of the probability 
distribution (see Table~\ref{tab:blr} and Fig.~\ref{fig:cnr_of}).  This 
deviation is moderately significant given our spectral resolution.   The main 
issue is that the systemic velocity has been derived from the symmetry of the 
rotation curve of the \OIII\ line that includes significant regions at radii 
$\gtrsim 3$\,kpc, where its shape is flat \citep{Perez1989AA} and is a 
reliable method.  The outflow model therefore requires that the black hole is 
offset by $\sim 380\,\mathrm{km\,s^{-1}}$ from the expected velocity.  While 
black hole recoil in merging systems has been discussed extensively in the 
literature and a significant minority are expected to have recoil speeds 
exceeding 500\,km\,s$^{-1}$ \citep{schnittman07}, there have been few convincing 
cases for such candidates \citep{komossa08,eracleous12,Komossa2012AdAst}.  
Moreover, this particular case would require a remarkable coincidence that the 
black hole recoil velocity exactly matches the outflow velocity of the BLR 
clouds.  This is another argument against the outflow model. 

We can try to avert this problem by fixing 
$\lambda_\mathrm{emit} = 2.2896$~\micron\ so it exactly matches the systemic 
velocity. Doing this results in a similar configuration to that in 
Fig.~\ref{fig:blr_view}, except that the weighting of the blue clouds in the 
near side is pushed to its lower limit at $\kappa \approx -0.5$.  And although 
the profile of the line wing is still matched reasonably well, the line core can 
no longer be fit as well as in Fig~\ref{fig:blr_of}\,(b).  Thus, although the 
flexibility of the P14 model means it is still able to fit the interferometric 
data well, the original rationale for trying the outflow model, with its 
increased number of parameters, is lost. A similar situation occurred when we
fixed $\xi=1$ so that we avoid the mid-plane transparency problem. The line profile then becomes
completely symmetric just as in the Keplerian model and again the rationale for using the
outflow model is lost.

Taking all these arguments together, we strongly favour the 
results of the Keplerian model as the most likely for IRAS~09149$-$6206, and
caution against over-interpreting the best-fit parameters of the outflow model.  
Nevertheless, we emphasise that the main results --- the photocentre offset and 
gradient, the BLR radius, and the black hole mass --- inferred from the 
Keplerian model and the outflow model are statistically consistent.  Despite 
these similarities in the key parameters, the two models are expected to show 
different characteristic features in velocity-resolved RM data 
\citep{Peterson2014SSRv}, and it would certainly be interesting to compare our 
results with dynamical modelling of high quality RM data for IRAS~09149$-$6206.

\begin{figure}
\centering
\includegraphics[width=0.4\textwidth]{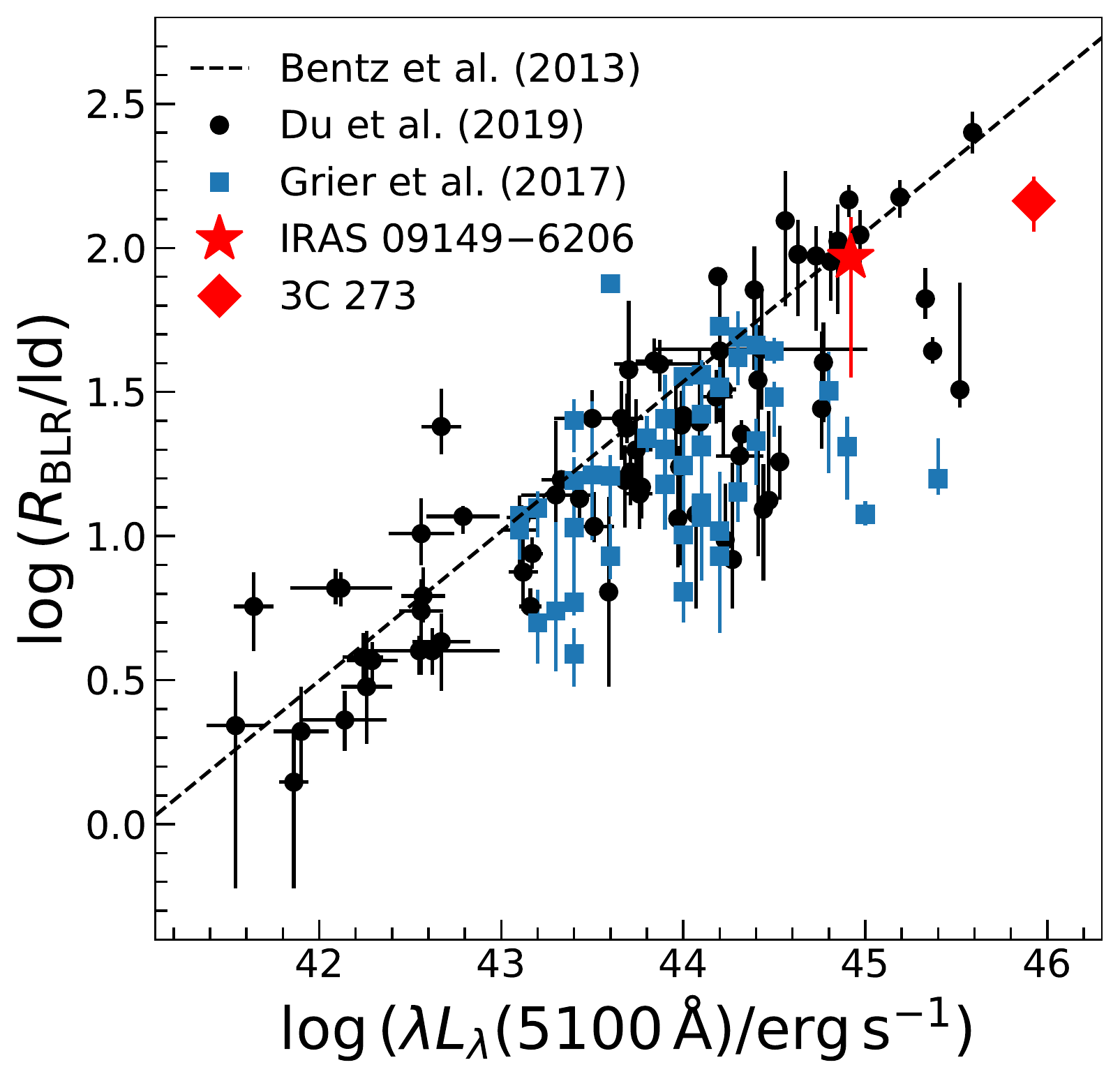}
\caption{Comparison of the BLR radii for IRAS~09149$-$6206 and 3C~273 measured 
with GRAVITY to those based on the reverberation mapping time lag.  The black 
circles are the RM measurements collected from the literature by 
\cite{Du2019ApJ}, which mainly include data from \cite{Bentz2013ApJ} and the 
SEAMBH campaign (see text for details).  The blue squares are based on H$\beta$ 
time lags from the SDSS-RM project \cite{Grier2017ApJa}.  The dashed line is the 
best-fit radius--luminosity relation from \cite{Bentz2013ApJ}.}
\label{fig:rl}
\end{figure}

\section{Black hole mass and the radius-luminosity relation} 
\label{sec:bhmass}

One of the parameters in the BLR model is the black hole mass.  For the 
Keplerian model, Table~\ref{tab:blr} shows that the best fitting value is 
$1.1\times10^8$\,M$_\odot$.  The posterior probability distributions in 
Fig.~\ref{fig:cnr_kp} show that it is correlated with the mean radius 
$R_{\rm BLR}$ and the inclination angle $i$, and this is likely what drives the 
large formal uncertainty of +0.3/$-$0.4\,dex.  Interestingly, 
Fig.~\ref{fig:cnr_of} shows that the correlation is weaker for the outflow 
model, and that the black hole mass derived is rather similar at 
$0.5\times10^8$\,M$_\odot$.  This is likely because the outflow velocities are 
linked to the local Keplerian speed in the P14 model.  As a result, as noted 
previously, one should be cautious when interpreting the black hole mass derived 
from the outflow model.

The only other report of the black hole mass is by \cite{Koss2017ApJ} for 
single-epoch estimates (which can have large uncertainties) based on the broad 
Balmer lines . Using the method of \cite{trakhtenbrot12} with the H$\beta$ line, 
these authors estimate a mass of $3.8\times10^{8}\,M_\odot$. Similarly, using 
the method of \cite{greene05} for the H$\alpha$ line they find a mass of 
$2.4\times10^{8}\,M_\odot$.  Although slightly less than these, our new value is 
fully consistent within the uncertainties of the methods used.  There is no 
reported measurement of the stellar velocity dispersion in the literature, 
making it difficult to place this object on the $M_{\rm BH}$--$\sigma_*$ 
relation.  However, as has been commonly done, we can use the width of the 
\OIII\ line. From the numbers reported by \cite{Perez1989AA}, we can estimate 
the dispersion as 250\,km\,s$^{-1}$.  This puts IRAS~09149$-$6206 only a factor 
2--3 below the relation as defined by \cite{gultekin09}, which is within the 
scatter.  Although the object is a factor 5 below the full relation of 
\cite{mcconnell13}, this offset is reduced when one considers the relation they 
find for late-type galaxies.  Good agreement is also found if we adopt the 
correlations for the late-type galaxy sample from \cite{Greene2019arXiv}.  This 
object therefore does not appear to be unusual in terms of its black hole mass.

We calculated the virial factor, 
$f_\mathrm{FWHM} \equiv G M_\mathrm{BH} / (R_\mathrm{BLR} v_\mathrm{FWHM}^2)$, 
by randomly drawing the BLR model parameters from the posterior parameter space
sampled from our fitting procedure.  The FWHM of the model line profile was used 
to calculate the velocity, 
$v_\mathrm{FWHM}$.  From the Keplerian model, 
$f_\mathrm{FWHM}=0.59^{+0.67}_{-0.26}$, while
$f_\mathrm{FWHM}=0.31^{+0.25}_{-0.09}$ from the outflow model, again 
statistically consistent with each other.  The typical $f_\mathrm{FHWM}$, based 
on calibration against the $M_{\rm BH}$--$\sigma_*$ relation, is $\sim 1.3$ 
(e.g., \citealt{Onken2004ApJ,Woo2010ApJ,Ho2014ApJ}).  The difference in the 
virial factor explains most of the difference between our BH mass and those from 
the single-epoch estimate.  Remarkably, \cite{Ho2014ApJ} found 
$f_\mathrm{FHWM} \approx 0.5$ for AGNs with pseudobulges.  Unfortunately, high 
resolution imaging is not available to reveal the bulge properties of 
IRAS~09149$-$6206.  Although the average $f_\mathrm{FHWM}$ from BLRs with 
dynamical modelling is about 1 \citep{Williams2018ApJ}, the inferred value of 
$f_\mathrm{FHWM}$ for individual AGNs shows a wide distribution.  In particular, 
the AGNs with similar BLR structure from RM dynamical modelling, Arp~151 
\citep{Pancoast2014MNRASb} and Mrk~142 \citep{Li2018ApJ}, also show a comparably 
low $f_\mathrm{FHWM}$.

The BLR size of IRAS~09149$-$6206 is robustly measured from our data.  In 
Fig.~\ref{fig:rl}, we compare the BLR radius of IRAS~09149$-$6206 and 3C~273 
measured with GRAVITY to the radius--luminosity (R--L) relation of the RM 
results.  \cite{Du2019ApJ} provide the most recent collection of 75 H$\beta$ 
time lags from various reverberation campaigns during the past two 
decades.\footnote{See Table~1 of \cite{Du2019ApJ} for more details.}  The 
collection primarily includes 41 AGNs from \cite{Bentz2013ApJ} and 25 AGNs with 
high accretion rate from the SEAMBH campaign 
\citep{Du2014ApJ,Du2015ApJ,Du2016ApJa,Du2018ApJb}.  We also compare with 44 
H$\beta$ time lags from the SDSS-RM campaign \citep{Grier2017ApJa}, which is 
based on a homogeneously selected quasar sample \citep{Shen2015ApJS}.  The BLR 
radius and 5100 \AA\ continuum luminosity of IRAS~09149$-$6206 are 89~light days 
and $8.32 \times 10^{44}\,\mathrm{erg\,s^{-1}}$ \citep{Koss2017ApJ}, 
respectively.  Those of 3C~273 are 146~light days (GC18) and 
$8.43 \times 10^{45}\,\mathrm{erg\,s^{-1}}$ \citep{Zhang2019ApJ}, respectively.  
Both IRAS~09149$-$6206 and 3C~273 show good consistency with the RM results.  
Moreover, IRAS~09149$-$6206 is very close to the best-fit relation from 
\cite{Bentz2013ApJ}.

Towards the high luminosity end of the R--L relation, the RM results also show 
large scatter and tend to drop below the best-fit relation from 
\cite{Bentz2013ApJ}, which we refer to as the `standard' R--L relation.  A study 
of high accretion rate AGNs finds a deviation from this relation that is 
primarily driven by accretion rate \citep{Du2015ApJ,Du2019ApJ}.  These authors 
proposed that the size of the BLR is reduced at high accretion rate due to the 
anisotropic emission of the `slim' accretion disk \citep{Abramowicz1988ApJ}.  
Interestingly, the BLR radii measured from the SDSS-RM campaign also lie mostly 
below the R--L relation, although the accretion rates of these AGNs are not high 
by the standard of the SEAMBH AGNs \citep{Grier2017ApJa,FonsecaAlvarez2019}.  
With the bolometric luminosity $\sim2 \times 10^{45}\,\mathrm{erg\,s^{-1}}$ 
(GC20a), the Eddington ratio of IRAS~09149$-$6206 is $\sim 0.2$.  Following 
Equation~(2) of \cite{Du2015ApJ} and adopting $i \approx 21\degree$ from our 
best-fit Keplerian model, the dimensionless accretion rate of IRAS~09149$-$6206 
is $\dot{\mathcal{M}} \approx 4.1$.  This is about the limit beyond which AGN 
start to deviate from the R--L relation.  With $i \approx 12\degree$ (GC18), the 
dimensionless accretion rate of 3C~273 is $\sim 23.8$.  The accretion rate 
estimates are quantitatively consistent with IRAS~09149$-$6206 being close to 
the R--L relation, while 3C~273 slightly deviates from it.

In closing, we note that different emission lines may trace regions at different 
radii of the BLR (e.g., 
\citealt{Gaskell1986ApJ,Clavel1991ApJ,Peterson1991ApJ,Kaspi2000ApJ,Bentz2010ApJ,
Grier2017ApJa}).  For example, the H$\alpha$ lag is expected to be larger than 
the H$\beta$ lag due to radial stratification and optical depth effects (e.g., 
\citealt{Rees1989ApJ,Korista2000ApJ,Korista2004ApJ,Bottorff2002ApJ,
Netzer2020MNRAS}).  \cite{Zhang2019ApJ} found the H$\beta$ time lag of 3C~273 is 
fully consistent with the BLR radius measured by GRAVITY from the Pa$\alpha$ 
emission, after de-trending the contamination from the jet emission 
\citep{Li2019arXiv}.  Considering that H$\beta$ and Pa$\alpha$ both come from 
the $n=4$ level of hydrogen, \cite{Wang2020NatAs} argued that the two lines are 
likely to originate from similar regions of the BLR.  They also estimated the 
possible size difference ($\sim 13\%$) of the H$\beta$ and Pa$\alpha$ emission 
regions based on the difference of their FWHM.  The size difference of H$\beta$ 
and Br$\gamma$ for IRAS~09149$-$6206 is less clear than that of 3C~273, as the 
two lines come from different upper levels.  However, we find the FWHM of 
Br$\gamma$ is $\sim 3350\,\mathrm{km\,s^{-1}}$, which is very close to the 
H$\beta$ FWHM $\sim 3500\,\mathrm{km\,s^{-1}}$ \citep{Perez1989AA}.  Following 
\cite{Wang2020NatAs}, we estimate a size difference between the H$\beta$ and 
Br$\gamma$ emitting regions of $\lesssim 10\%$.

\section{Origin of the spatial offset between BLR and continuum photocentre} 
\label{sec:offset}

\begin{figure}
\centering
\includegraphics[width=0.4\textwidth]{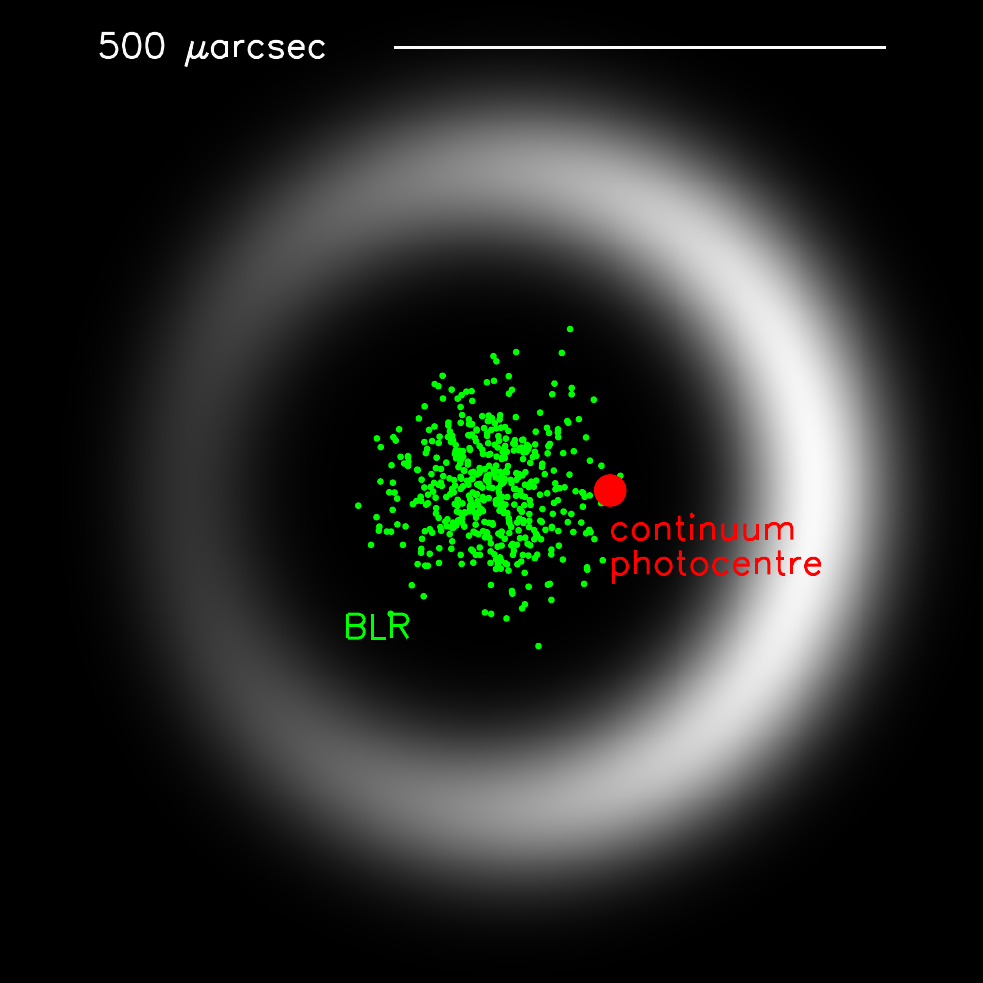}
\caption{Cartoon illustrating a possible cause for the observed offset between 
the near-infrared continuum photocentre and the BLR. Dust continuum is assumed 
to follow a ring that is centred on the BLR and with radius based on the size 
measurement of GC20a. Brightness variation along this ring causes the 
offset between BLR and continuum photocentre. Variation in brightness of the hot 
dust distribution along a ring is supported by the resolved observations of 
NGC~1068 reported by \cite{GCngc1068}.}
\label{fig:cartoon}
\end{figure}

Our models (Table~\ref{tab:blr}) place the continuum photocentre outside the 
bulk of the BLR cloud distribution, at a physical scale 
$\sim 0.14$~pc.  While there could be many possible explanations for 
such an offset, we argue that it is consistent with the simplest explanation: 
both BLR and dust are centred on the black hole but there is a modest level of 
asymmetry in the $K$-band emission which is arising from hot dust on scales 
larger than the BLR, near the sublimation radius.  
An asymmetry might arise from differential brightness between the near and far 
sides, from clumpiness or irregularities in the emitting dust structure, or 
possibly also if the edge of a foreground dust lane crosses the line of sight to 
the nucleus.  The first of these is illustrated in the cartoon of 
Fig.~\ref{fig:cartoon}, where, inspired by the asymmetric ring-like dust 
emission of NGC~1068 \citep{GCngc1068} we assume a dust ring with radius based 
on the $K$-band size measurement for IRAS~09149$-$6206 by GC20a.  A linear 
brightness gradient is imposed to reproduce the shift between geometric centre 
of both BLR and dust ring, and the continuum photocentre.  Such asymmetries will 
lead to nonzero continuum closure phases, but these remain small for compact 
sources. For the specific configuration of Fig.~\ref{fig:cartoon}, closure 
phases $< 1$\degree\ are expected, consistent with the measurements shown in 
Fig.~\ref{fig:t3}.  More generally, in the marginally resolved limit, the 
maximum closure phase on VLTI triangles is approximately, 
$1^\circ (\rm FWHM/0.6 \rm mas)^3$, where the source FWHM is scaled to that 
measured from FT data for IRAS 09149-6206.

The Fig.~\ref{fig:cartoon} scenario is clearly not unique, but one of the simple 
and plausible ways to create an offset between BLR and continuum photocentre, 
while staying consistent with all GRAVITY observations. Other plausible ways 
include a tilted view at non-planar (e.g. bowl-shaped) dust emission.  And 
finally, more exotic explanations for an offset are not ruled out, such as the 
recoil option discussed in Sec.~\ref{sec:compmod}.

\section{Conclusion}
\label{sec:conc}

With 7.8 hours on-source integration of GRAVITY, we successfully spatially 
resolve the broad Br$\gamma$ emission line region of IRAS~09149$-$6206.  This is 
the second source, following 3C~273 (GC18), for which near-infrared 
interferometric observations directly constrain the size of the BLR and enable 
an estimate of the mass of the central black hole.  With an improved phase 
calibration method, the differential phase can be uniformly calibrated to 
systematic uncertainty $\sim 0.05\degree$ for each baseline.  This enables us to 
robustly resolve the BLR of the nearby AGN with the broad Br$\gamma$ line.  The 
main results are summarised as follows.

\begin{itemize}
\item We obtain a $\sim 0.5\degree$ differential phase signal on two baselines, 
which is measured from the Br$\gamma$ emission line with a peak flux that is 
$\sim 6\%$ of the continuum.  The differential visibility amplitude of the BLR 
is $\sim 0.8\%$ above the continuum, indicating that the BLR is much more 
compact than the continuum emission.  The closure phase of the continuum 
emission is $\sim 0\degree$, consistent with the continuum being only 
marginally resolved. 
\item The model-independent reconstruction of photocentres reveals that the BLR 
is offset to the east of the photocentre of the continuum by $\sim 120\,\mu$as.  
While the offset dominates the differential phase signal, the 
photocentres display a significant blue--red velocity gradient in a north--south 
direction, indicating that we are resolving the kinematics of broad Br$\gamma$ 
emission. 
\item We model the interferometric data with (1) a simplified BLR model 
including only clouds on circular orbits and (2) a generalised 
dynamical model that allows for radial motions, and which is widely used in 
analysing AGN reverberation mapping data.  Both models provide an adequate fit 
to the data.  We argue against the outflow model because there are several 
difficulties associated with its physical interpretation and implication, and 
caution is needed when interpreting the parameters from the fit.  Based on the 
favoured Keplerian model, and with 95\% credibility intervals, we report a 
radius for the BLR of $65_{-39}^{+30}\,\mu$as or $89_{-53}^{+41}$ light days, 
and a black hole mass $1.1_{-0.84}^{+1.80} \times 10^8\,M_\odot$.
\item The BLR radii measured by GRAVITY (Br$\gamma$ size for IRAS~09149$-$6206 
and Pa$\alpha$ size for 3C~273) are quantitatively consistent with the 
radius--luminosity relation based on H$\beta$ reverberation mapping of AGNs.
\end{itemize}


\begin{acknowledgements}
We thank the referees for their careful reading of the manuscript and 
their suggestions that have helped to improve it.  This research has made use 
of the NASA/IPAC Extragalactic Database (NED), which is operated by the Jet 
Propulsion Laboratory, California Institute of Technology, under contract with 
the National Aeronautics and Space Administration. J.D. was supported in part by 
NSF grant AST 1909711.  A.A. and P.G. were supported by Funda\c{c}\~{a}o 
para a Ci\^{e}ncia e a Tecnologia, with grants reference UIDB/00099/2020 and 
SFRH/BSAB/142940/2018.  SH acknowledges support from the European Research 
Council via Starting Grant ERC-StG-677117 DUST-IN-THE-WIND.  J.S. thanks the 
helpful discussions with Jian-Min Wang, Pu Du, Yan-Rong Li, and Yu-Yang 
Songsheng.
\end{acknowledgements}


\bibliography{blr}{}
\bibliographystyle{aa}


\appendix

\section{Calibration of the pipeline reduced differential phase}
\label{apd:phase}

\begin{figure*}
\centering
\includegraphics[width=0.7\textwidth]{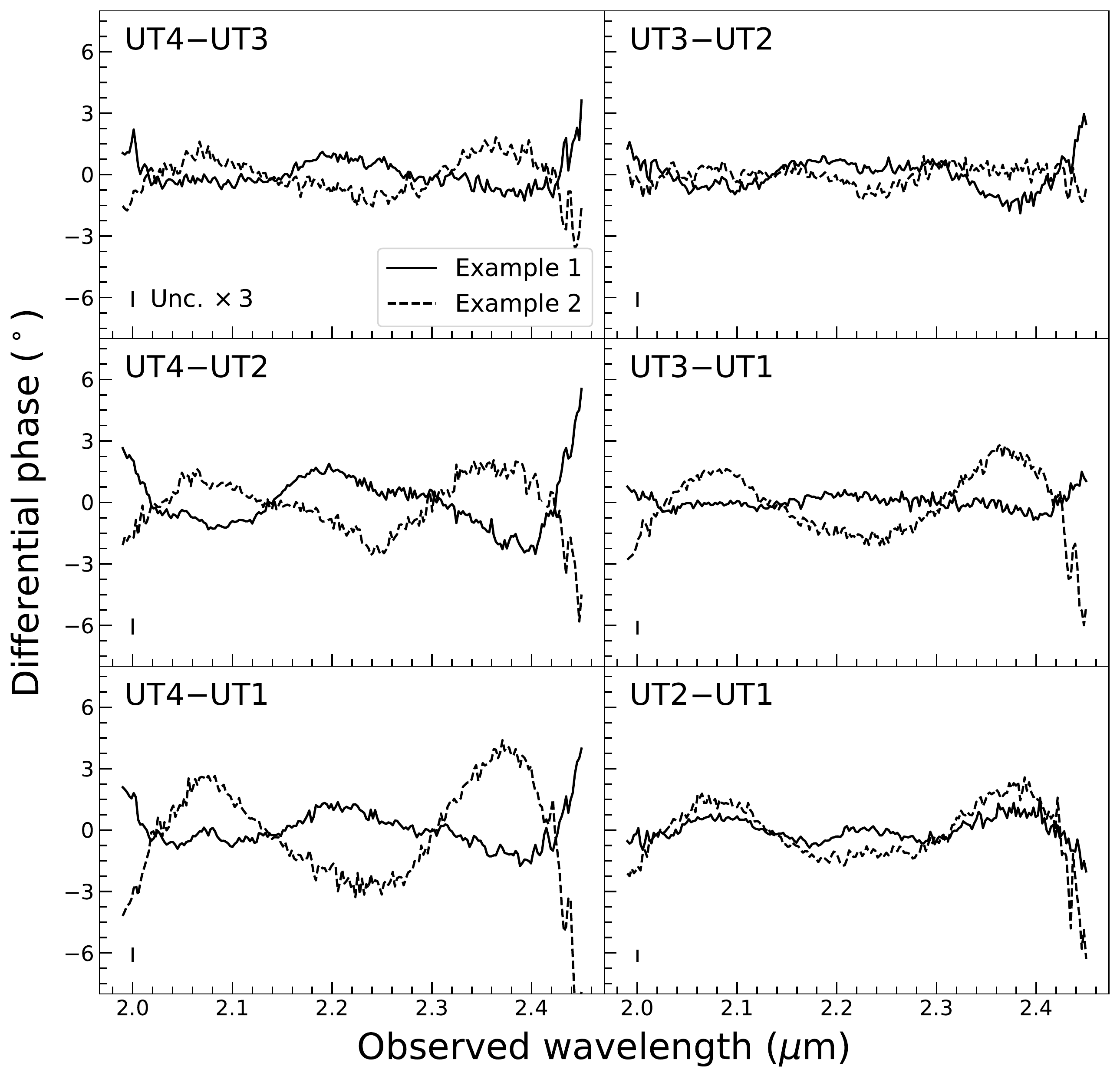}
\caption{Variation of the differential phase of GRAVITY data.  
The solid and dashed curves display the differential phases of 
two observations of calibrator stars, whose differential phase should be 
consistent with $0\degree$.  The vertical bar in the lower left corner of each 
panel indicates the typical uncertainty, enlarged by a factor of three for 
clarity.}
\label{fig:phivar}
\end{figure*}

\begin{figure}
\centering
\includegraphics[width=0.4\textwidth]{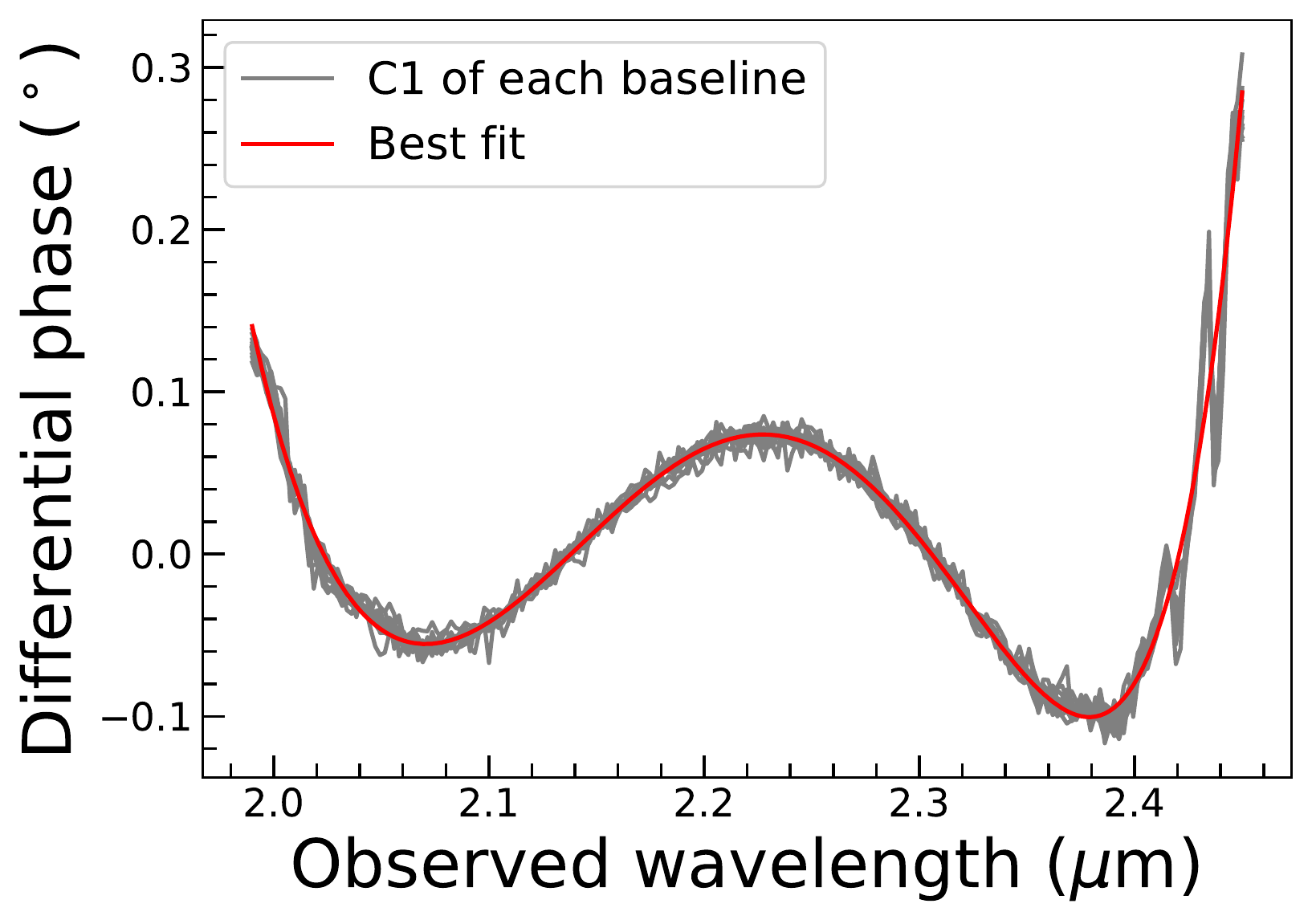}
\caption{First principle components (C1) from the PCA of all of the 
baselines (grey curves), before and after the GRAVITY intervention, 
separately, are very consistent with each other.  We fit them together with 
an 8th order polynomial (red curve).  The strong variations beyond 2.4 \micron\ 
are due to the resonances of water vapour \citep{Colavita2004PASP}.}
\label{fig:c1}
\end{figure}

\begin{figure*}
\centering
\includegraphics[width=0.7\textwidth]{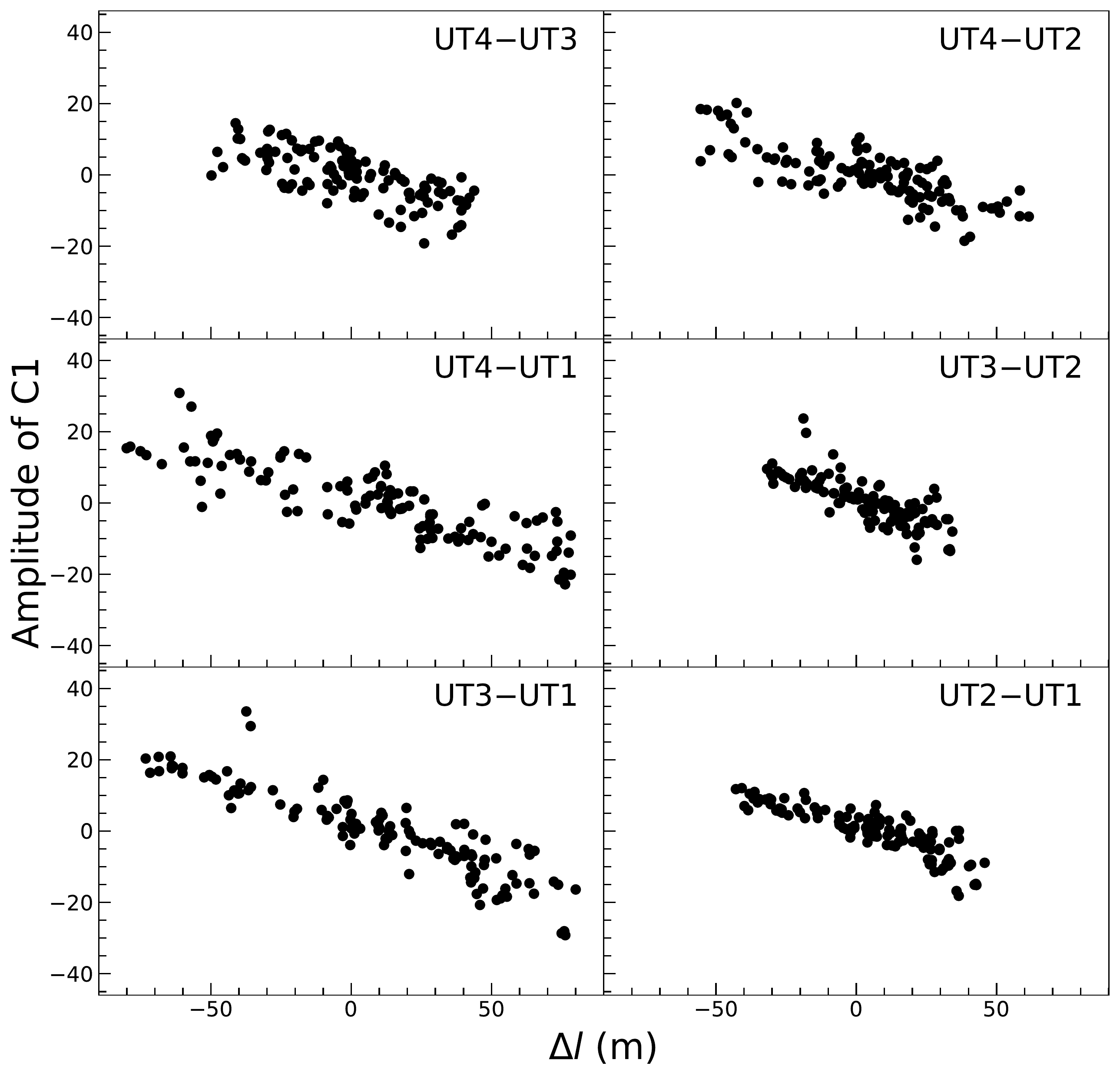}
\caption{Amplitude of the C1 component as a function of the length 
difference between the non-vacuum delay lines in each baseline.}
\label{fig:air}
\end{figure*}

\begin{figure*}
\centering
\includegraphics[width=0.7\textwidth]{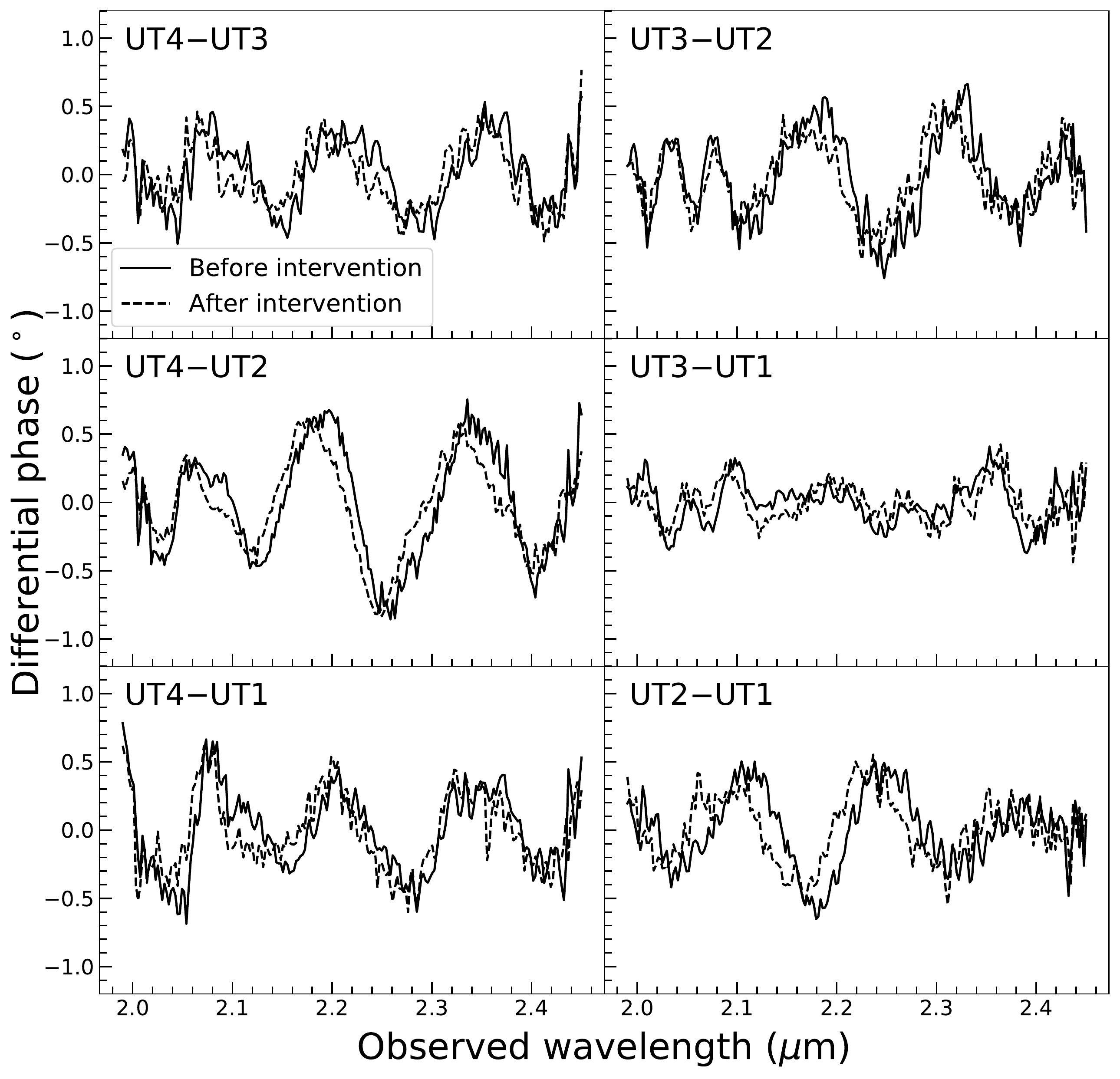}
\caption{Stable phase features for each baseline, before (solid) and after 
(dashed) the intervention of GRAVITY.}
\label{fig:phimean}
\end{figure*}

\begin{figure*}
\centering
\includegraphics[width=0.7\textwidth]{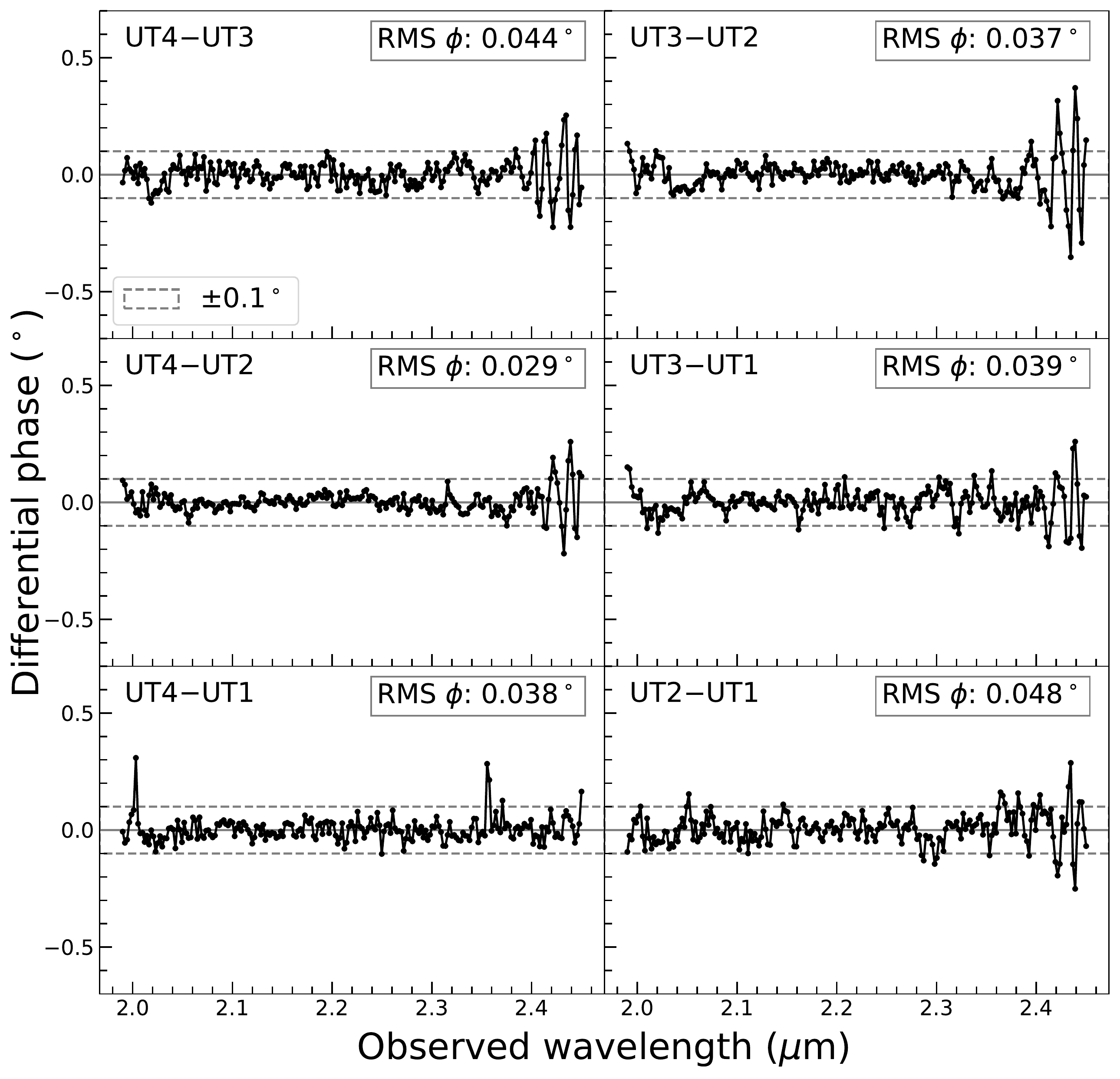}
\caption{Differential phase of the calibrator stars, flattened and stacked 
in the same way as the AGN.  The 23 calibrator files after the GRAVITY 
intervention in late 2019 are used, which is quantitatively close to the results 
using the 122 files before the intervention.  The RMS of the phase at 2.05 to 
2.40 $\mu$m is reported on the upper right corner of each panel.}
\label{fig:calib}
\end{figure*}

As shown in Fig.~\ref{fig:phivar}, the pipeline reduced differential phases of 
the calibrator stars show, at different times, considerable non-zero variation 
as a function of wavelength.  To investigate this, we have collected from the 
ESO archive\footnote{\url{http://archive.eso.org/wdb/wdb/eso/gravity/form}} 
on-axis observations of calibrator stars with the same spectral resolution and 
polarisation configuration. Before the exchange of the GRAVITY science detector 
during the intervention in October 2019, we found 122 good files.\footnote{We 
require the fringe tracking ratio of the data better than 99.9\%.} We found 23 
good files after the intervention, up to March 2020.  The calibrator data reveal 
a clear systematic phase variation.  A principle component analysis\footnote{We 
adopted \texttt{sklearn.decomposition.PCA} function \citep{scikit-learn}.} (PCA) 
shows that there is only one variable component dominating the whole variation 
of the differential phase in all of the baselines.  We refer to this variable 
component as C1 (Fig.~\ref{fig:c1}), which stays the same both before and after 
the intervention of GRAVITY.

Using PCA, we also find the variability is introduced by the dispersion of the 
air in the non-vacuum delay line.  The amplitude of the C1 component is well 
correlated with the light path difference in the delay line 
(Fig.~\ref{fig:air}).  In addition, we also find stable instrumental phase 
features, that are likely to come from the Fabry-P\'{e}rot effect of the 
elements in the cryostat of GRAVITY.  As shown in Fig.~\ref{fig:phimean}, these 
features changed after the GRAVITY intervention, which is expected since the 
cryostat was opened and many elements were adjusted.

We therefore flatten the pipeline reduced AGN data by fitting and subtracting an 
instrumental phase model. This includes the stable phase feature as well as the C1 
component, the amplitude of which is determined from the fit.  The data before and 
after the intervention are fit separately using the appropriate stable phase 
feature.  We measure the systematic uncertainty of the differential phase after 
removing the instrumental features using the calibrator data.  As 
Fig.~\ref{fig:calib} shows, the calibrator differential phases, observed after 
the intervention, are reduced and stacked in the same way as the AGN data.  The 
stacked differential phases are fully consistent with zero, and have an RMS 
close to, or better than, 0.05\degree.  The phase at $\lesssim 2.05\,\micron$ and 
$\gtrsim 2.40\,\micron$ suffers strongly from the absorption bands of carbon 
dioxide and water in the atmosphere, therefore the phase variation increases at 
those wavelengths.  For calibrator data taken before the intervention, we find 
the RMS of the stacked differential phases is quantitatively similar to that in  
Fig.~\ref{fig:calib}.  We therefore conclude that the systematic uncertainty 
of the differential phase is 0.05\degree.  We also find that the accuracy of the 
flattening is insensitive to the choice of the wavelength mask (e.g. across an 
emission line), as long as the mask is only a reasonable fraction of the entire 
spectrum.  This is because we only have one free parameter in the fit, the 
amplitude of the C1 component, which has an unvarying profile.

The pipeline uses a 3rd order polynomial to fit the phase of the complex 
visibility ($R + iI$) of each DIT to derive the self-reference phase 
($\phi_\mathrm{self}$).  The corrected visibility 
($R^\prime + i I^\prime$) of each DIT is,
\begin{eqnarray}
R^\prime =& \cos(\phi_\mathrm{self}) R - \sin(\phi_\mathrm{self}) I, \nonumber\\ 
I^\prime =& \sin(\phi_\mathrm{self}) R + \cos(\phi_\mathrm{self}) I
\end{eqnarray}
It would be very convenient if we could avoid the \textit{a posteriori} 
flattening by adopting a better self-reference phase model that takes into 
account the dispersion of the air.  Unfortunately, we find that the models of 
the refractive index of the air are not accurate enough, in terms of their 
functional form, for our accuracy requirement \citep{Ciddor1996ApOpt}.  We also 
tested whether a higher order polynomial fit can remove the effect of the 
dispersion of the air.  However, even fitting with a 7th order polynomial still 
leaves a residual larger than 0.1\degree, and using such a high order polynomial 
is risky because it may also fit the scientific phase signal.  Therefore, we 
prefer to apply the separate phase flattening method described above after the 
default pipeline data reduction.

\section{The differential phase of AGN BLRs}
\label{apd:visphase}

In this section, we derive the generalised differential phase of the AGN including 
the continuum phase, under the assumption that the BLR and continuum are 
marginally resolved.  The complex visibility is defined as,
\begin{equation}\label{eq:visdef}
\tilde{V} = V(\lambda) \, e^{i\phi_\lambda} = 
\frac{\iint I_\lambda(\vec{\sigma})\,e^{-2\pi i\, \vec{u} \cdot \vec{\sigma}} \,dl\,dm}
{\iint I_\lambda(\vec{\sigma}) \,dl\,dm},
\end{equation}
where $V(\lambda)$ and $\phi_\lambda$ denote the visibility amplitude and phase 
as functions of wavelength ($\lambda$); $I_\lambda(\vec{\sigma})$ is the source 
intensity distribution; $\vec{\sigma}=(l,m)$ coordinate on the sky; 
$\vec{u}=\vec{B}/\lambda=(u,v)$ is the baseline vector.  When the source is only 
marginally resolved, i.e., $2\pi\,\vec{u} \cdot \vec{\sigma} \ll 1$, we have
$e^{-2\pi i\, \vec{u} \cdot \vec{\sigma}} \approx 1 - 2\pi i\, \vec{u} \cdot 
\vec{\sigma}$ and $e^{i\phi_\lambda} \approx 1 + i \phi_\lambda$.  Therefore, 
from equation~(\ref{eq:visdef}) and setting $V(\lambda) \approx 1$,
\begin{equation} \label{eq:vphi}
\phi_\lambda \approx -2\pi\,\vec{u} \cdot \iint I_\lambda(\vec{\sigma}) 
\vec{\sigma} \,dl\,dm = -2\pi\,\vec{u} \cdot \vec{x}_\lambda,
\end{equation}
where $\vec{x}_\lambda$ is defined as the photocentre of the source at the 
wavelength $\lambda$.  The phase is proportional to the photocentre of the 
source projected onto the baseline.  For three baselines that form a closed 
triangle, i.e., $\vec{u}_1 + \vec{u}_2 + \vec{u}_3=0$, their closure phase is
naturally,
$\phi_{1,\lambda}+\phi_{2,\lambda}+\phi_{3,\lambda}=-2\pi\,
(\vec{u}_1 + \vec{u}_2 + \vec{u}_3) \cdot \vec{x}_\lambda = 0$.

The observed AGN emission consists of two components, the continuum emission from 
the hot dust and the line emission from the BLR.  Therefore, the observed 
complex visibility is,
\begin{equation}
\tilde{V} = \frac{f_{c,\lambda} \tilde{V}_c + f_{\mathrm{BLR},\lambda} 
\tilde{V}_\mathrm{BLR}}{f_{c,\lambda} + f_{\mathrm{BLR},\lambda}},
\end{equation}
where $f_{c,\lambda}$ and $\tilde{V}_c$ are the spectral flux and complex 
visibility of the continuum emission as functions of the wavelength, while 
$f_{\mathrm{BLR},\lambda}$ and $\tilde{V}_\mathrm{BLR}$ are the flux and complex 
visibility of the BLR.  Under the marginally resolved assumption, the 
differential phase is
\begin{equation}\label{eq:dphi}
\Delta \phi_\lambda = \phi_\lambda - \phi_c = 
\frac{f_\lambda}{1 + f_\lambda}\, (\phi_{\mathrm{BLR},\lambda} - \phi_c),
\end{equation}
where $\phi_c$ is the phase of the continuum, which is not expected to vary with 
different spectral channels; $\phi_{\mathrm{BLR},\lambda}$ is the phase of 
the BLR as a function of wavelength; and 
$f_\lambda \equiv f_{\mathrm{BLR},\lambda}/f_{c,\lambda}$ is the emission line 
flux normalised by the continuum. Taking Equation~(\ref{eq:vphi}) for the 
continuum and BLR separately into Equation~(\ref{eq:dphi}), we obtain
\begin{equation} \label{eq:dphot}
\Delta \phi_\lambda = -2\pi\, \frac{f_\lambda}{1 + f_\lambda}\, 
\vec{u} \cdot (\vec{x}_\mathrm{BLR,\lambda} - \vec{x}_c),
\end{equation}
where $\vec{x}_\mathrm{BLR,\lambda}$ and $\vec{x}_c$ are the photocentres of 
the BLR and the continuum.  

When we use the BLR model to fit the data, we calculate the differential phase 
based on Equation~(\ref{eq:dphi}).  The BLR clouds contribute to 
$\phi_\mathrm{BLR}$, while the offset of the BLR contributes to the continuum
phase,
\begin{equation} \label{eq:dphi_fit}
\Delta \phi_\lambda = \frac{f_\lambda}{1 + f_\lambda}\, [\phi_{\mathrm{BLR},\lambda} 
- 2\pi\,(u x_\mathrm{o} + v y_\mathrm{o})],
\end{equation}
where $\{x_\mathrm{o}, y_\mathrm{o}\}$ is the vector of the offset of the BLR 
with respect to the photocentre of the continuum.

\section{Bayesian inference}
\label{apd:bayes}

We infer the optimal model parameters ($\vec{\Theta}$) based on the 
interferometric data ($D$) and our prior knowledge of the source and the model 
($I$), according to Bayes' theorem,
\begin{equation}
p(\vec{\Theta} | D, I) = \frac{p(\vec{\Theta} | I) p(D | \vec{\Theta}, I)}{p(D | I)},
\end{equation}
where $p(\vec{\Theta} | D, I)$ is the posterior probability density function of the 
model parameters.  The prior, $p(\vec{\Theta} | I)$, is provided by our prior 
knowledge about the probability distribution of the model parameters.  The 
evidence, $p(D | I)$, is useful to compare across different models using the 
Bayes factor, 
\begin{equation}
K=\frac{p(D | I_\mathrm{model~2})}{p(D | I_\mathrm{model~1})},
\end{equation}
assuming the prior knowledge of model 1 and model 2 are equivalent.  The 
likelihood function, $\mathcal{L} = p(D|\vec{\Theta},I)$ is defined assuming a 
Gaussian probability distribution,
\begin{eqnarray}
\ln\,\mathcal{L} = -\frac{1}{2} \sum_i^n 
\left(\frac{(f_i - \tilde{f}_i(\vec{\Theta}))^2}{\sigma_{f,i}^2} + \ln\,(2\pi 
\sigma_{f,i}^2)\right) &&\\
-\frac{1}{2} \sum_i^n 
\left(\frac{(\phi_i - \tilde{\phi}_i(\vec{\Theta}))^2}{\sigma_{\phi,i}^2} + 
\ln\,(2\pi \sigma_{\phi,i}^2)\right)&,&
\end{eqnarray}
where $f_i$ and $\sigma_{f,i}$ are the normalised line flux and its uncertainty, 
respectively, at the $i$th spectral channel; $\phi_i$ and $\sigma_{\phi,i}$ are 
the differential phase and its uncertainty, respectively;\footnote{For 
simplicity, and with reference to Equation~(\ref{eq:dphi}), we write $\phi$ 
rather than $\Delta\phi$ here.} and $\tilde{f}_i(\vec{\Theta})$ and 
$\tilde{\phi}_i(\vec{\Theta})$ are the normalised line flux and differential 
phase from the BLR model.

We sample the posterior distribution and calculate the Bayesian evidence for 
each of the two models using the nested sampling method implemented in the 
Python package, \texttt{dynesty} \citep{Speagle2020MNRAS}.  The nested sampling 
method has been extensively used for reverberation mapping studies of the BLR 
(e.g., \citealt{Pancoast2014MNRASb,Grier2017ApJb,Li2018MNRAS,Li2018ApJ,
Li2019arXiv,Williams2018ApJ,Raimundo2019MNRAS,Raimundo2020MNRAS,Wang2020NatAs}), 
due to its efficiency in sampling the posterior even when there are complex 
multimodal structures.  The best-fit model parameters and the uncertainties 
(Table~\ref{tab:blr}) are estimated, respectively, from the maximum 
{\it a posteriori} value and the 95\% (2 $\sigma$) credible interval of the 
posterior probability density distributions.  The Bayesian evidences are used to 
compare the probability of the two models with the Bayes' factor.  We also 
calculate the $\chi_\mathrm{r}^2$, Bayesian information criterion 
(BIC, \citealt{Schwarz1978AnSta}), and Akaike information criterion 
(AIC, \citealt{Akaike1973,Hurvich1989}),
\begin{eqnarray}
\chi_\mathrm{r}^2 &=& \Sigma^N_i \left( \frac{(f_i - \tilde{f}_i(\vec{\Theta}))^2}{\sigma_{f,i}^2} + \frac{(\phi_i - \tilde{\phi}_i(\vec{\Theta}))^2}{\sigma_{\phi,i}^2} \right) \bigg/ (N - k), \\
\mathrm{BIC} &=& k \ln\,N - 2\ln\,\mathcal{L}_\mathrm{max}, \\
\mathrm{AIC} &=& 2k - 2\ln\,\mathcal{L}_\mathrm{max} + \frac{2k (k + 1)}{N - k - 1}
\end{eqnarray}
where $N$ is the number of data points, $k$ is the number of free parameters, 
and $\mathcal{L}_\mathrm{max}$ is the maximum likelihood based on the samples 
from the fit.

\section{Plots of the BLR model fits and results}
\label{apd:fit}

\begin{figure*}
\centering
\includegraphics[width=0.9\textwidth]{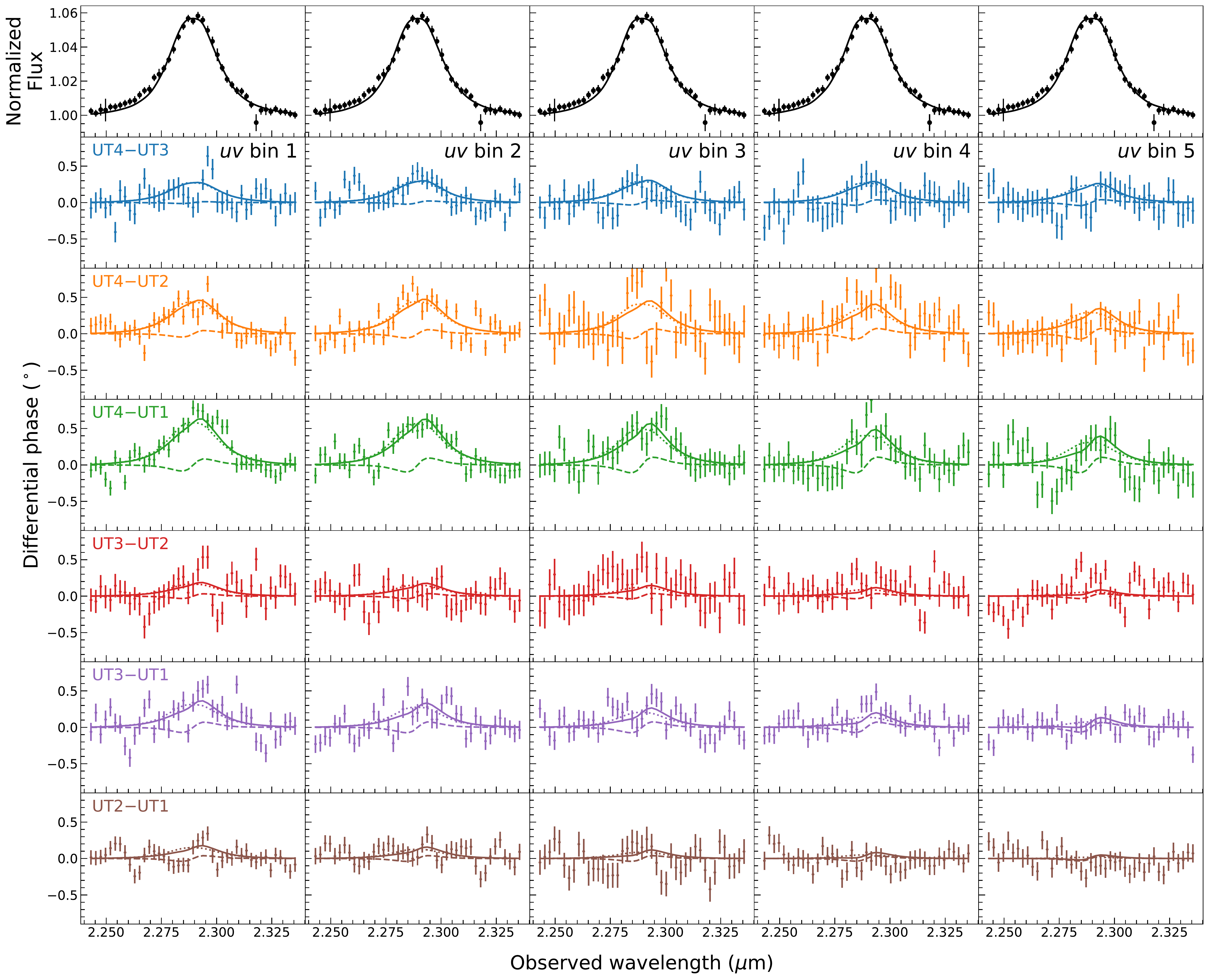}
\caption{The Keplerian model best-fit spectrum and phases.  The first row 
displays the same normalised spectrum.  The black solid line is the best-fit 
model.  The rows are the differential phases of different baselines.  Each 
column displays the phases of each $uv$ bin.  In each panel, the solid line is 
the total phase signal, which sums up the differential phase of the BLR (dashed 
line) and the continuum phase (dotted line) due to the offset between the BLR 
and the photocentre of the continuum.}
\label{fig:phs_kp}
\end{figure*}

\begin{figure*}
\centering
\includegraphics[width=0.9\textwidth]{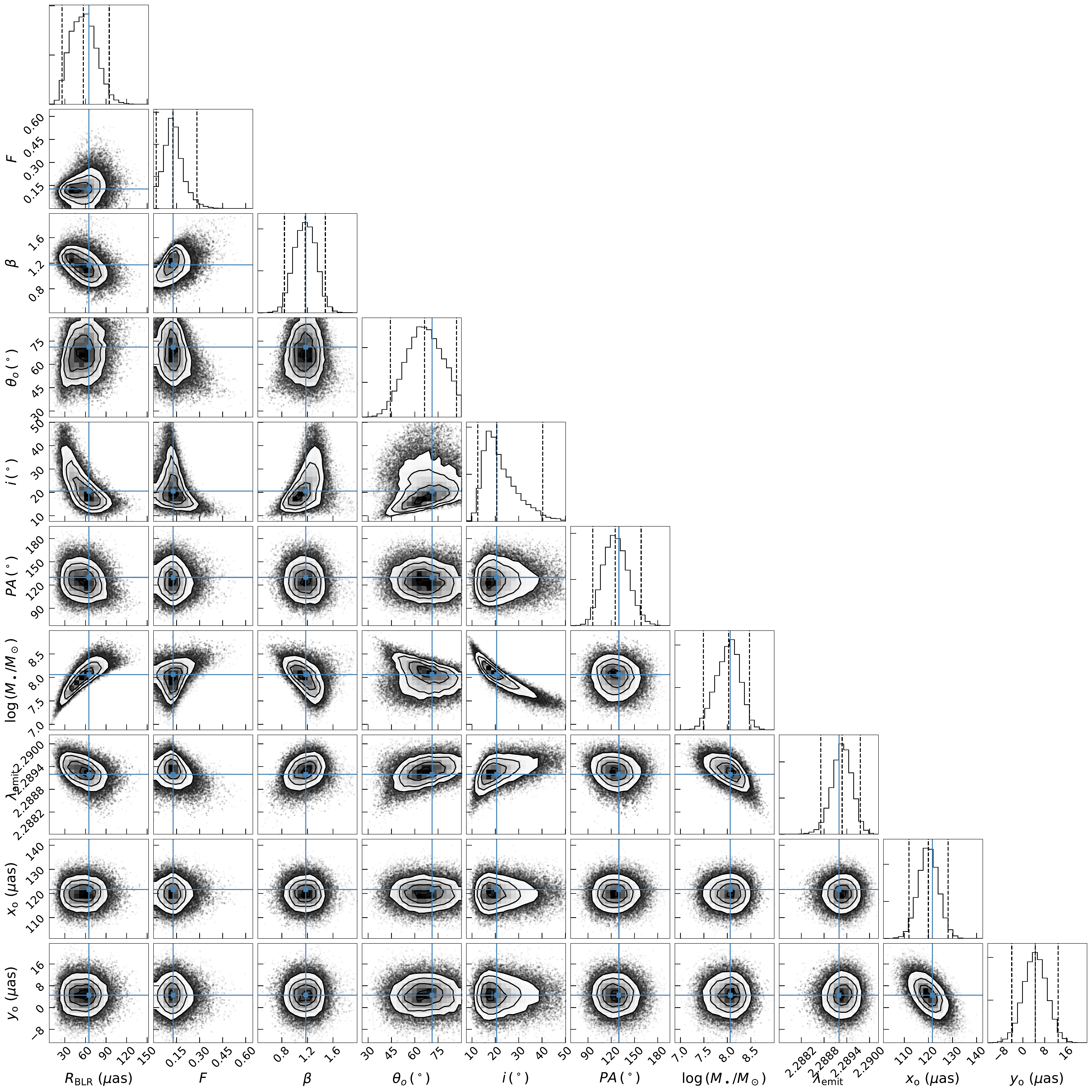}
\caption{The posterior probability distribution of selected parameters from 
the Keplerian model.  The blue lines and the cross points represent the maximum 
a posteriori.}
\label{fig:cnr_kp}
\end{figure*}

\begin{figure*}
\centering
\includegraphics[width=0.9\textwidth]{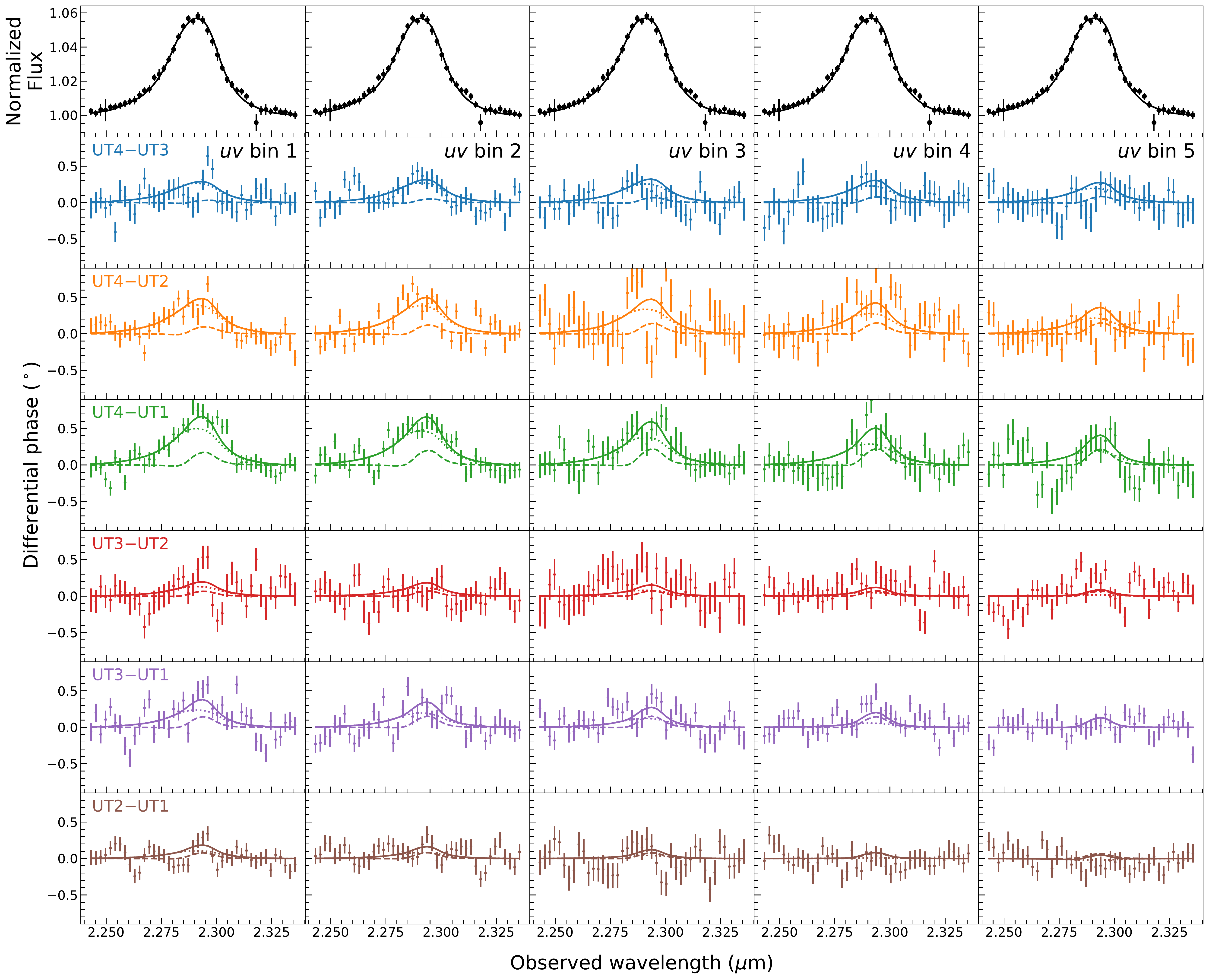}
\caption{The outflow model best-fit spectrum and phases.  The arrangements and 
symbols are the same as Fig.~\ref{fig:phs_kp}.}
\label{fig:phs_of}
\end{figure*}

\begin{figure*}
\centering
\includegraphics[width=0.9\textwidth]{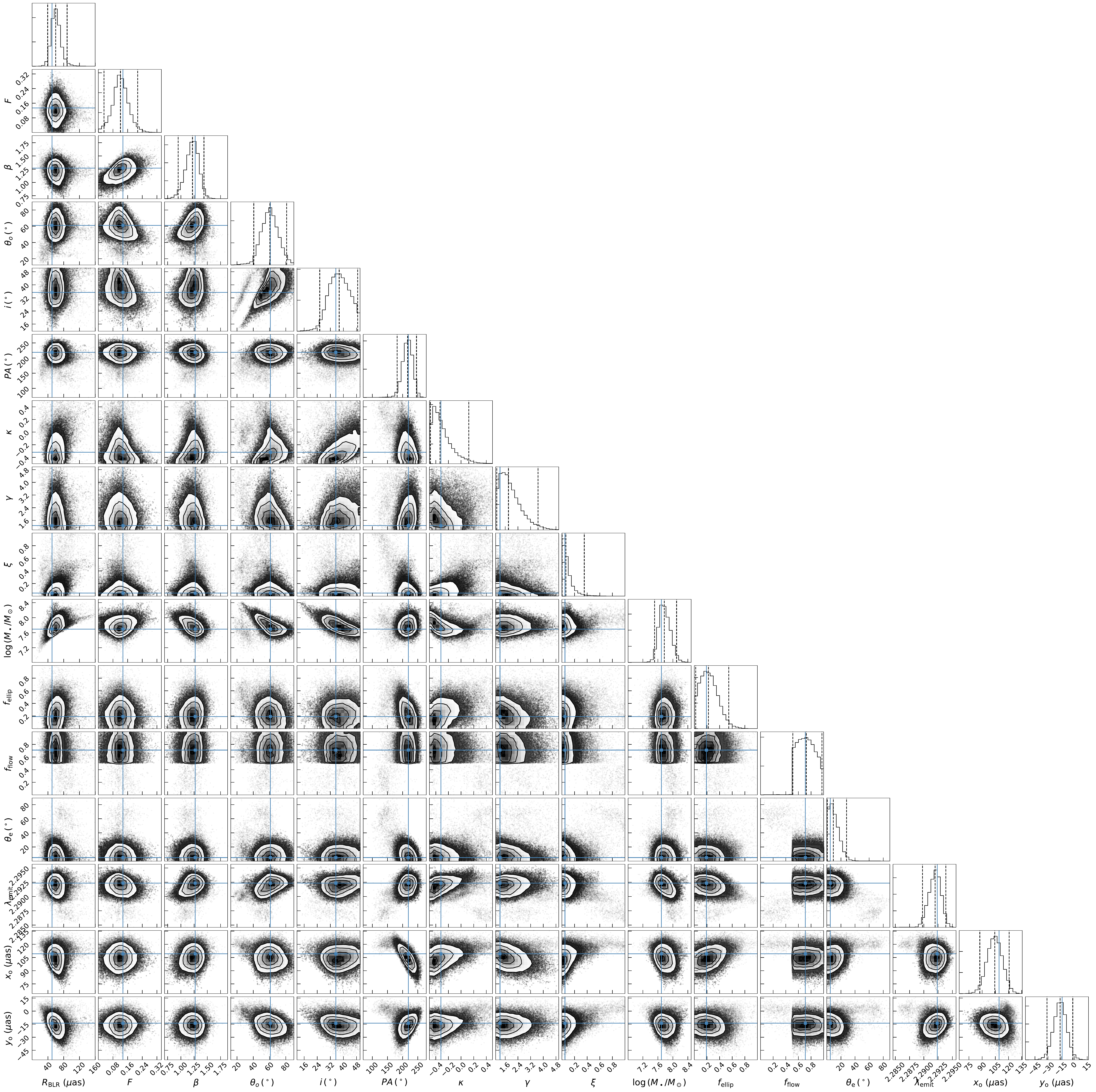}
\caption{The posterior probability distribution of selected parameters from 
the outflow model.  The blue lines and the cross points represent the 
maximum a posteriori.}
\label{fig:cnr_of}
\end{figure*}

The fits to the flux spectra and the differential phases for each baseline and 
angular {\it uv} bin are displayed in Fig.~\ref{fig:phs_kp} and 
Fig.~\ref{fig:phs_of} for the Keplerian model and the outflow model 
respectively.  The models were fit to the 5 bins of averaged differential phase 
based on $uv$ coverage (Fig.~\ref{fig:uvc}) in order to minimise smearing and 
ensuring the signal is still visible.  The posterior probability density 
distributions of the physical parameters that are important for our discussion 
are plotted in Fig.~\ref{fig:cnr_kp} and Fig.~\ref{fig:cnr_of}.

\section{Additional tests of the BLR modelling}
\label{apd:test}

In order to reduce the number of parameters, we considered a simplified 
version of the P14 model in which we fix 
$\sigma_{\rho,\mathrm{circ}}=\sigma_{\Theta,\mathrm{circ}}=0$, so that bound 
orbits are always circular, and 
$\sigma_{\rho,\mathrm{radial}}=\sigma_{\Theta,\mathrm{radial}}=0$ so that orbits 
with radial motion are always at the same location on the phase space ellipse 
defined by Equation~(\ref{eq:ellip}).  In addition we set 
$\sigma_\mathrm{turb}=0$ so that there is no additional macroturbulence.  The 
outcome is that the inferred parameters of this simplified model are entirely 
consistent with the full P14 model.  This and further tests suggest that these 
five technical parameters ($\sigma_{\rho,\mathrm{circ}}$, 
$\sigma_{\Theta,\mathrm{circ}}$, $\sigma_{\rho,\mathrm{radial}}$, 
$\sigma_{\Theta,\mathrm{radial}}$, and $\sigma_\mathrm{turb}$) are less 
important in terms of defining the model.  This is consistent with the fact that 
these parameters are largely ignored by \cite{Pancoast2014MNRASb} and other 
works that use the model to fit RM data.  In the main text here, we therefore 
also do not discuss the specific values of these technical parameters.

We interpret the dominant differential phase signal as the continuum phase, 
which is produced by the offset between the BLR and the photocentre of the 
continuum emission.  However, in Section~\ref{sec:outmod} on the outflow model, 
we show that the BLR alone is also able to produce an asymmetric differential 
phase with an all positive signal shape.  It is therefore interesting to test 
whether the outflow model is able to fit the data without the need for a 
continuum phase, so that the BLR would lie at the centre of the continuum 
emission. We are indeed able to fit the differential phase data reasonably well.  
However, the inferred radius of the BLR is very large, with 
$R_\mathrm{BLR} \approx 200\,\mu\mathrm{as}$.  This is easily understood because 
the best-fit dynamical model in Section~\ref{ssec:blrmod} has a mean radius 
$R_\mathrm{BLR} \approx 60\,\mu\mathrm{as}$, which corresponds to a phase signal 
$\sim 0.15\degree$ shown in Fig.~\ref{fig:blr_of}\,(b).  When the continuum phase 
is fixed to be zero, the BLR size has to increase by a factor of $\sim 3$ in 
order to fit the much larger $\sim 0.5\degree$ phase signal.  However, in GC20a 
we found that the size of the continuum emission for IRAS~09149$-$6206, measured 
as a Gaussian FWHM, is only 0.54--0.64 mas.  Thus its BLR radius should be  
$<100$ $\mu$as, since it must be significantly smaller than the continuum.  
Similarly, following the method presented in Sec.~\ref{sec:blrvamp}, we find the 
differential visibility amplitude of the large BLR model to be about a factor 2 
lower than the other two models shown in Fig.~\ref{fig:vamp}.  Therefore, the 
differential visibility amplitude data strongly disfavour the large BLR model.

We also attempted to include the differential visibility amplitude data in the 
fit.  However, as the differential visibility amplitude is very sensitive to the 
relative size between BLR and the continuum, the inferred BLR size strongly 
depends on the assumed FWHM of the continuum emission.  Using 
$\mathrm{FWHM} = 0.6$~mas, our inferred BLR sizes are only $21_{-9}^{+22}$ and 
$30_{-18}^{+24}$ $\mu$as for the Keplerian and outflow models respectively.  
GC20a estimated the FWHM of the continuum emission to be $0.54 \pm 0.05$ mas 
by directly fitting the visibility amplitude of the fringe tracker data (the 
`FT size' in their Table~2).  From the differential visibility amplitude of the 
science channel, they also obtained a size of $0.64 \pm 0.06$~$\mu$as 
(the `SC size').  Under the marginally resolved limit, the latter method only 
yields the correct continuum size when the BLR is exactly a point source; for 
better resolved sources, the derived quantity is the quadrature difference 
between the continuum emission size and the BLR size \citep{Waisberg2017ApJ}.  
Therefore, we should expect the SC size to be slightly smaller than the FT size.  
This is true for 3C~273 and PDS~456 but not for IRAS~09149$-$6206.  Although the 
problem can be easily explained by the uncertainty, this indicates that the 
continuum size of IRAS~09149$-$6206 is still quite uncertain.  Therefore, we 
exclude the differential visibility amplitude data from our primary BLR fittings 
and use it only as a consistency check.

For our last test, we check whether the input parameters can be recovered 
by fitting mock data generated by the Keplerian model (mock~1) and the outflow 
model (mock~2).  We also test whether the Bayes factor, BIC, and AIC provide a 
reliable judgement on the model fits.  We use the best-fit parameters of the 
Keplerian model and the outflow model to generate mock data, adopting Gaussian 
noise based on the measured uncertainty.  In the next step, we fit both sets of 
mock data with both the Keplerian model and the full P14 model.  The input 
parameters are usually recovered if the same model is used to generate and to 
fit the mock data.  Focusing on the mock~2 data, the best-fit Keplerian model 
results are similar to those in Fig.~\ref{fig:phs_kp} and Fig.~\ref{fig:blr_kp}.  
The averaged phase, after subtracting the best-fit continuum phase, still shows 
clear S-shape profiles.  To compare the two models fitting the mock~2 data, the 
Bayes factor and AIC prefer the P14 model, while the BIC incorrectly selects the 
Keplerian model.  For the fits to mock~1 data, the P14 model can naturally fit 
the data and provide the parameters consistent with the input, which is as 
expected.  The Bayes factor determines the Keplerian model and the outflow model 
are equivalent.  The AIC incorrectly prefers the outflow model, while the BIC 
prefers the Keplerian model.  In a nutshell, these tests demonstrate that: 
(1) our models infer the parameters self-consistently when the mock data are 
generated from them and (2) The Bayes factor provides a reasonable judgement on 
the preferred model.

\end{document}